\documentclass{ar-1col}
\usepackage[numbers,sort&compress]{natbib}

\usepackage{graphicx,amssymb,amsmath}
\usepackage{color,ulem}
\usepackage{bm,soul}

\renewcommand{\Re}{\mathop \mathrm{Re}}

\newcommand{\rev}[1]{{\color{black}#1}}
\jname{Annu. Rev. Condens. Matter Phys.}
\jvol{10}
\jyear{2019}
\begin{document}
% Page header
\markboth{J. P. Pekola and I. M. Khaymovich}{Thermodynamics in Single-Electron Circuits and Superconducting Qubits}

% Title
\title{Thermodynamics in Single-Electron Circuits and Superconducting Qubits}

%Authors, affiliations address.
\author{J. P. Pekola$^1$ and I. M. Khaymovich$^{2,3}$
\affil{$^1$QTF Centre of Excellence, Department of Applied Physics, Aalto~University~School of Science, P.O. Box 13500, 00076 Aalto, Finland;
email:~jukka.pekola@aalto.fi}
\affil{$^2$Max Planck Institute for the Physics of Complex Systems, N\"othnitzer Stra{\ss}e~38, 01187 Dresden, Germany;
email:~ivan.khaymovich@pks.mpg.de}
\affil{$^3$Institute for Physics of Microstructures, Russian Academy of Sciences, 603950~Nizhny Novgorod, GSP-105, Russia}
}

\date{\today}

\begin{abstract}
Classical and quantum electronic circuits provide
ideal platforms to investigate stochastic thermodynamics and they have served as
a stepping stone to realize Maxwell's demons with highly controllable
protocols. In this article we first review the central thermal phenomena in quantum nanostructures.
Thermometry and basic refrigeration methods will be described as enabling
tools for thermodynamics experiments. Next we discuss the
role of information in thermodynamics which leads to the concept of
Maxwell's demon. Various Maxwell's demons realized in single-electron circuits
over the past couple of years will be described. Currently true quantum
thermodynamics in superconducting circuits is in focus of attention, and
we end the review by discussing the ideas and first experiments in this
exciting area of research.
%We review the central thermal phenomena in quantum nanostructures. Thermometry and basic refrigeration methods will be described as enabling tools for thermodynamics experiments. Next we discuss the role of information in thermodynamics which leads to the concept of Maxwell's demon. Classical and quantum electronic circuits provide ideal platforms to investigate stochastic thermodynamics that serves as a stepping stone to realize Maxwell's demons with highly controllable protocols. Various Maxwell's demons realized in single-electron circuits over the past couple of years will be described. Currently true quantum thermodynamics in superconducting circuits is in focus of attention, and we end the review by discussing the ideas and first experiments in this exciting area of research.
\end{abstract}

%Keywords, etc.
\maketitle

%Table of Contents
\tableofcontents

\section{INTRODUCTION}

\subsection{Thermal properties of quantum nanostructures}\label{Sec:Calorimetry+Bolometry}
Thermodynamics of small systems, where fluctuations play a key role, is a topic of intense experimental interest currently. Theoretical framework has been laid over the past decades \cite{bochkov1977general,bochkov1981nonlinear,Evans1993,Gallavotti1995,Jarzynski1997,Crooks1999,Seifert2005,Seifert2012_RepProgPhys,Sagawa2008,Sagawa2010}, but first realizations in laboratory have come up much later \cite{Toyabe2010, Saira2012, Koski2013, Berut2012, Orlov2012, Jun2014, Koski2014_PRL, Koski2014_PNAS, Roldan2014, Koski2015, Khaymovich2015, Pekola2015NatPhys, Hong2016, Vidrighin2016, Gavrilov2017, Ribezzi,Collin2005,Alemany2011,Alemany2015}. In this article we present experiments on electronic circuits, which present highly favorable systems for the experimental studies of stochastic thermodynamics. As compared to soft-matter systems, where fluctuation relations were first investigated experimentally, electronic circuits present certain advantages. First of all, the circuits are stable meaning that experiments can be repeated many times under essentially identical conditions: this allows for large statistics, normally up to $10^6$ experimental realizations can be achieved as compared to typical $10^2\ldots10^3$ repetitions in soft-matter experiments. Secondly, the Hamiltonian of the circuits is usually very simple, and the system is governed by it accurately; moreover electrons can be monitored one by one and coupling to heat baths can also be modeled precisely. Finally, the experiments are usually performed at low temperatures, which means that circuits made out of superconducting metals behave quantum mechanically: they thus present an ideal testbed to investigate thermodynamics of {\sl open quantum systems}.

Our main interest in most
of this review is the electron system in the nano-structures. We deal primarily with conductors formed of ordinary metals, where even the smallest structures (volume of $\sim 10^{-22}$ m$^3$ can be realized by standard electron-beam lithography based nano-fabrication) contain about $10^{8}$ conduction electrons. This large number means that it is possible to assign a well defined temperature to such a conductor of electrons, if this subsystem is in local equilibrium. It is quite fortunate that this is indeed the case in most experiments at sub-kelvin temperatures, like the ones presented in this review. First of all, under these conditions the electrons are effectively thermally isolated from the phonon bath and locally from other conductors typically either by superconducting leads or tunnel barriers. Another important point is that even in clean elemental metals, the electron-electron scattering rate $\gamma_{\rm e-e}$ is as large as $10^9$ s$^{-1}$. This rate is larger than a typical injection rate $\gamma$ of non-equilibrium carriers in the experiments that we discuss. For instance $\gamma \sim 1 $ s$^{-1}$ in the Szilard's Engine \cite{Koski2014_PRL,Koski2014_PNAS}, and $\gamma \sim 10^6$ s$^{-1}$ in the Autonomous Maxwell's demon (MD) experiments \cite{Koski2015}. The local temperature of the electrons $T_e$ can then differ from that of the phonon bath $T$ because the electron-phonon relaxation rate $\gamma_{\rm ep}$ scales as $T^{3}$ in ordinary metals, and assumes a value of about $\gamma_{\rm ep} \sim 10^{5}$ s$^{-1}$ at a typical operating temperature of $T=0.1$ K. Putting all above together means that electrons follow Fermi-Dirac distribution where their temperature $T_e$ is then determined by external biasing conditions and coupling to other conductors.

\begin{marginnote}[120pt]
\entry{MD}{Maxwell's demon}
\end{marginnote}

\begin{figure}
\includegraphics[width=0.9\textwidth]{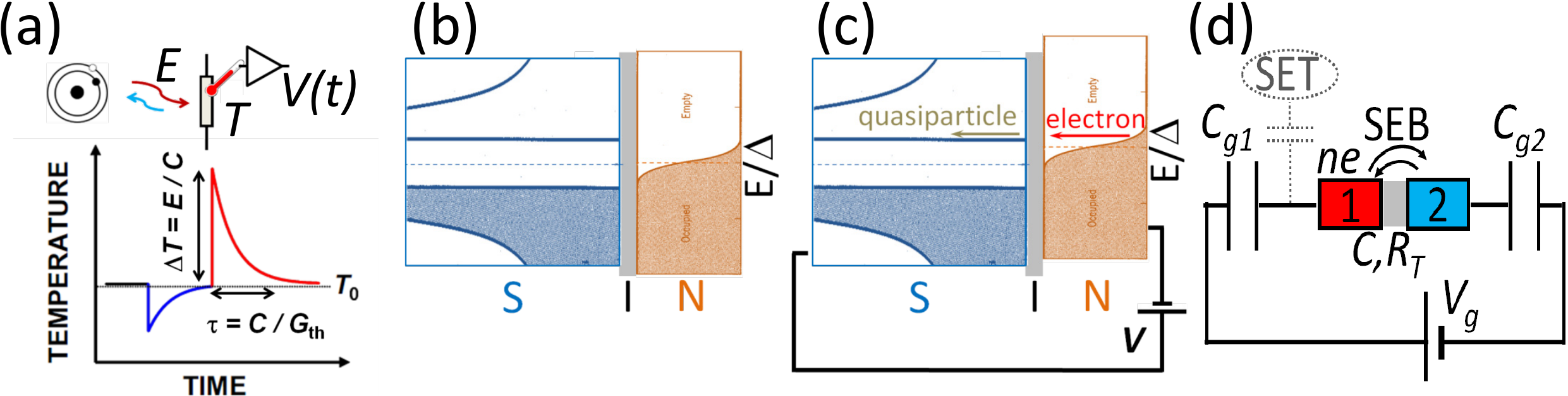}
\caption{
(a)~Schematics of calorimeter setup. Upper panel shows a normal metal resistor absorbing photons from an ``artificial atom'' (qubit etc.), temperature of which is measured in real time via voltage $V(t)$ (see the bottom panel).
(b)~Energy diagram of the NIS junction at a temperature $T$.
The energies are considered with respect to the Fermi energy of the normal metal normalized to the superconducting gap.
Solid lines show the density of states in the normal metal $n_N(E)$ and superconductor $n_S(E)$, occupation numbers are described by filled areas.
(c)~The same NIS energy diagram as in (b), but under a finite voltage bias $V$ demonstrating refrigeration principle.
Electrons tunnel from the normal metal at energies above the Fermi level to the superconductor, cooling the former and heating up the latter.
(d)~The schematics of the single-electron box (SEB) based on two Coulomb-blockaded islands ($1$ and $2$) tunnel coupled to each other (grey area) and voltage biased $V_g$ with respect to each other via capacitive couplings $C_{g1,2}$ to the leads.
The tunnel junction is characterized by the resistance $R_T$ and the capacitance $C$.
The electron tunneling between islands is shown by black arrows together with the charge $ne$ of the left island.
The charge detector shown by grey dashed lines is formed by a single-electron transistor (SET).
}
\label{Fig:Calorimetry+NIS+SEB}
\end{figure}

In the scenario illustrated above, one can then distinguish two regimes of operation in terms of response to external injection of heat. The most common and straightforward scheme corresponds to quasi-stationary power $\dot Q$, which results in a steady-state temperature change $\delta T_e$ with respect to that under equilibrium conditions. In this kind of ``bolometric'' detection, the signal, i.e. the temperature change, depends on the thermal conductance $G_{\rm th}$ to the bath, which is most often given by the electron-phonon coupling. We have for small changes of temperature, $\delta T_e / T_e \ll 1$, $\delta T_e =\dot Q /G_{\rm th}$. It is worth noting the basic but often overlooked feature of bolometric detection: the steady state temperature does not depend on the heat capacity $C_e$ of the electron system. On the contrary, the temporal evolution of $T_e$ is governed by the heat capacity and $G_{\rm th}$. In the particular case of abrupt absorption of energy $Q$ in the electron system, the local temperature changes - again the linear regime is assumed - instantaneously (within the time $\sim \gamma_{\rm e-e}^{-1}$) by the amount $\delta T_e = Q/C_e$. Following this transient, the local temperature then returns back to the equilibrium value exponentially with the relaxation time $\tau =C_e/G_{\rm th}$ as $\delta T_e(t)= \frac{Q}{C_e} e^{-t/\tau}$, see Fig.~\ref{Fig:Calorimetry+NIS+SEB}(a). Detecting such energy quanta thermally, ``calorimetry'', will be described later in Sec.~\ref{Sec:Calorimetry_single_photon_detection} of this review. One more ingredient to add to this simplified picture of temperature variation of the electron sub-system is the issue of fundamental energy fluctuations. Due to the coupling to the bath, the energy of the electron system fluctuates around its mean value by $\langle \delta E^2\rangle = k_B T_e^2 C_e$. The frequency spectrum of these fluctuations extends up to $\tau^{-1}$. This noise can be facilitated for instance by the fluctuating impact of phonons interacting with electrons. Such fluctuating energy then translates into fluctuating effective temperature of the electron system with $\langle \delta T_e^2\rangle =k_B T_e^2 /C_e$, which sets the fundamental detection limit of temperature. In practical sense this temperature noise can be as large as $1$~mK at $T_e = 100$~mK for a small metallic conductor described in the beginning of this Section.
For electron-phonon limited heat transport, the fluctuations extend then up to the cut-off frequency $\gamma_{\rm e-p} \sim 10^5$~Hz. Measuring temperature in the DC and AC regimes by a few particular thermometers will be described in the next Section.

\begin{marginnote}[120pt]
\entry{SEB}{single-electron box}
\entry{SET}{single-electron transistor}
\entry{qubit}{quantum bit}
\end{marginnote}

The main focus of this review is in measurements of work and heat in small systems. The mesoscopic electronic systems at low temperatures provide a nearly unique set-up to detect small amounts of heat directly via temperature variations as described above. Heat is then the energy that is distributed to the electron gas. The detectability of temperature variations is a favorable consequence of the combination of the effective isolation of the system from the bath and the extremely small heat capacity. The heat capacity is determined by the nearly free Fermi gas at a temperature of $\sim 10^{-6}T_F$, where $T_F$ is the Fermi temperature. Work, on the other hand, is in these systems the change of the energy of the few degrees of freedom on which the external force is applied to. This can be given by the charging state of the single-electron circuit (electron number on the box) or the level populations in a quantum bit (qubit). We will discuss these issues in detail in Sections~\ref{Sec:W,Q,S_in_single-electronics} and \ref{Sec:ThermDyn_in_OQS}.
\subsection{Thermometry}\label{Sec:Thermometry}
Temperature of the electron system is coded in its energy distribution of the Fermi-Dirac form, i.e., $f(E)=1/(1+e^{E/k_B T_e})$, where $E$ is the energy of electron with reference at the Fermi level. It is therefore necessary to find sensitive probes of the distribution for thermometry. The probe should also lead to as little back-action as possible, that is it should not disturb the system by driving it into non-equilibrium.

Tunnel probes provide a convenient measurement of temperature. The rate of tunneling in a junction from electrode 1 to 2 typically through an insulating barrier is given by
\begin{gather}\label{e0_Gamma_rate}
\Gamma_{1\to 2}(\delta E)= \frac{1}{e^2R_T}\int dE n_1(E)n_2(E+\delta E) f_1(E)[1-f_2(E+\delta E)] \ .
\end{gather}
Here $R_T$ is the resistance of the tunnel contact, $n_i(E)$ are the normalized (with the corresponding normal state value) densities of states (DOS) and $f_i(E)$ the distributions in electrodes $i=1,2$. Here $\delta E$ is the energy gain in the tunneling event, which is determined by the voltage bias $eV$ in the absence of charging effects; for the case of Coulomb effects it will be discussed separately.
Further we assume an electron-hole symmetry to be satisfied for DOSes $n_i(-E) = n_i(E)$ and
distribution functions $f_i(-E) = 1-f_i(E)$ in both electrodes leading to $\Gamma_{1\to 2} (\delta E) = \Gamma_{2\to 1} (\delta E)$,
and omit the rate subscript where it does not cause a confusion.
\begin{marginnote}[120pt]
\entry{DOS}{density of states}
\entry{N}{normal metal}
\entry{I}{insulator}
\entry{S}{superconductor}
\end{marginnote}

For the sake of our arguments later on, let us take an NIS junction (N - normal metal, I - insulator, S - superconductor) where 1 is `S' with Bardeen-Cooper-Schrieffer DOS $n_S(E)=|\Re(E/\sqrt{E^2-\Delta^2})|$ and 2 is `N' with $n_2(E)=1$. Here $\Delta$ is the superconducting gap. Then for a voltage biased junction $\Gamma_{S\to N}(eV)= \frac{1}{e^2R_T}\int dE n_S(E) f_S(E)[1-f_N(E+eV)]$, with the electron charge $-e<0$. %The rate $\Gamma_{N->S}(\delta E)$ can be calculated similarly using \eqref{e0_Gamma_rate}.
The average charge current $I(V) = e \left(\Gamma_{S\to N}(eV) - \Gamma_{N\to S}(-eV)\right) = -I(-V)$ from N to S is then given in this system by
\begin{equation} \label{e1_I(V)_fN}
I(V)= \frac{1}{2eR_T}\int dE n_S(E) [f_N(E-eV)-f_N(E+eV)].
\end{equation}
For thermometry this equation implies that current through the NIS junction depends only on the temperature of N \cite{Rowell1976}, i.e., its distribution $f_N(E)$, since $n_S(E)$ is temperature independent for practical purposes when the temperature is much lower than the critical temperature $T_C$ of the superconductor (see Fig.~\ref{Fig:Calorimetry+NIS+SEB}(b)).
This makes an NIS junction a very favorable choice for thermometry, working down to few millikelvin range of temperatures \cite{Feshchenko2015}.
Current or conductance can be monitored in either steady-state conditions or, more recently, also to follow fast changes of temperature.

\subsection{Refrigeration}\label{Sec:Refrigeration}
Another important result for an NIS junction yields the average heat current $\dot Q(V)$ into the normal electrode N as
\begin{equation} \label{e2_Q(V)}
\dot Q(V)= \frac{1}{e^2R_T}\int dE (E-eV)n_S(E) [f_N(E-eV)-f_S(E)].
\end{equation}
The implications of this relation are that due to the gap $\Delta$ and the singularities in the DOS of S, the power is negative for $|eV|<\Delta$, i.e. the biased junction refrigerates N (and heats S), see Fig.~\ref{Fig:Calorimetry+NIS+SEB}(c), and for large biases $|eV| \gg \Delta$ it Joule heats both the N and S electrodes \cite{Nahum1994,Leivo1996,Clark2005,Giazotto2006}.
Both Eqs. \eqref{e1_I(V)_fN} and \eqref{e2_Q(V)} will be useful in what follows.

\section{%Single-electron circuits: work, heat, entropy and fluctuation relations
SINGLE-ELECTRON CIRCUITS: WORK, HEAT, ENTROPY AND FLUCTUATION RELATIONS}\label{Sec:W,Q,S_in_single-electronics}
\subsection{Single-electron box}
Up to now we have assumed tunneling with no Coulomb effects included. We noted earlier, however, that the energy gain in tunneling can be generalized to $\delta E$ instead of $eV$, which is essential when the additional energy of one electron counts.
Let us focus on a single-electron box (SEB), depicted in Fig.~\ref{Fig:Calorimetry+NIS+SEB}(d). It consists of a tunnel junction with capacitance $C$, where the rates are governed by the formulae presented above \eqref{e0_Gamma_rate}, and of a gate capacitance $C_g=C_{g1}+C_{g2}$ and voltage $V_g$ applied on it. We denote the total capacitance $C_\Sigma = C+C_g$.
Irrespective of the type of the electrodes of the tunnel contact, the electrostatic energy $U(n,n_g)$ of the SEB with $n$ extra electrons on it is that of a capacitor with
\footnote{This charging energy is modified due to the so-called parity effect when one or both islands are superconducting and the energy gap in the low-temperature limit shifts up the energy of the state with odd electrons.}
\begin{equation} \label{e3_U(n,ng)}
U(n,n_g)=E_C(n-n_g)^2.
\end{equation}
Here $E_C=e^2/2C_\Sigma$ is the ``charging energy'' and $n_g=C_gV_g/e$ is the polarization charge in units of $e$ induced by the gate voltage. The polarization charge $n_g$ is a continuous variable, whereas $n$ assumes naturally only integer values. (Yet $n$ can be both negative and positive, since as noted before, the total number of electrons is $\sim 10^8$ or larger, but the Coulomb energy fixes its value typically to within few adjacent values.)

\subsection{Work, heat and entropy}
The SEB can be viewed as a system where the work on it is done by the gate voltage source, and the bath is that formed of the electron gas (that is eventually coupled to the ``super-bath'' of phonons). It is now quite straightforward to write the expressions of work $W$ and heat $Q$ to the bath in a driven process. This is a realization specific quantity, since we are dealing with ``stochastic thermodynamics'', where standard thermodynamic results emerge as averages over many realizations. The heat is the irreversible part of the energy that is associated with inelastic transitions from the point of view of the Coulomb energy. For simplicity, and to represent a typical experimental situation, we limit to the situation where $n$ can assume values $0$ and $1$ only, when the gate parameter is within $0\le n_g\le 1$. This happens, when the temperature of the electrons is low enough, $k_B  T_e \ll E_C$. Based on Eq. \eqref{e3_U(n,ng)}, $Q_i$, the loss of energy to the bath in the transition occurring at gate position $n_{g,i}$, where electron tunnels into ($+$) or out from ($-$) the box is given by $Q_i= \pm E_C(2n_{g,i}-1)$. If there are several transitions within the measurement protocol, the total heat is then
\begin{equation} \label{e4_Q_sum_over_jumps}
Q= \sum_i \pm E_C(2n_{g,i}-1).
\end{equation}
To find the total work $W$ performed to the system, we need to add the change in internal energy of the system in the particular realization, i.e.
\begin{equation} \label{e5_1st_law}
W%-\Delta F
= \Delta U+Q,
\end{equation}
where $\Delta U = U(n^f,n_g^f)-U(n^i,n_g^i)$. Here superscripts $i,f$ refer to the initial and final configurations, respectively. For some experiments to be presented, the protocol runs from $n_g^i =0$ to $n_g^f =1$, and in this case the internal energy change simplifies into $\Delta U =(1-n^i-n^f)E_C$.
The dissipated work, $W_d\equiv W-\Delta F$, appearing in some fluctuation relations, is related to \eqref{e5_1st_law} through the equilibrium free energy difference $\Delta F = F(n_g^f)-F(n_g^i)$ between the end points of the drive, $F(n_g^k) = -k_B T\ln[\sum_n \exp(-U(n,n_g^k)/k_B T)]$, $k=i,f$.
In equilibrium state of the system, i.e., when $\Delta F = \Delta U - T \Delta S_{sys}$, the dissipated work defines the total entropy production on the trajectory
\begin{gather}\label{e6_S_tot_equil}
W_d/T = \Delta S \equiv \Delta S_{sys} + \Delta S_{bath}
\end{gather}
as the sum of system $\Delta S_{sys} = -\ln (P_{n^f}^f/P_{n^i}^i)$ and bath $\Delta S_{bath} = Q/T$ entropy productions.

In general one can
%One can also
assign the bath entropy production to each $i$th transition between $n_{i-1}$ and $n_i$ in an experiment as soon as the tunneling rates at the particular values of control parameter(s), i.e. gate voltage, are known
\begin{gather}\label{e7_S_bath}
\Delta S_{bath} = \sum_i \ln \frac{\Gamma_{n_{i-1}\to n_i}(n_{g,i})}{\Gamma_{n_{i}\to n_{i-1}}(n_{g,i})}
 \ .
\end{gather}
This expression summed with $\Delta S_{sys}$ is particularly useful to determine the total entropy production $\Delta S = \Delta S_{sys} + \Delta S_{bath}$ in the situations when the precise temperature of the bath is unknown or this reservoir is in non-equilibrium.
The system entropy $S_{sys} = -\ln P_n$ is still determined by the instantaneous probability distribution $P_n$.

The presented expressions allow us to discuss thermodynamics in SEB. From the point of view of primary value of the measurements, one should, however, keep in mind that measuring the transitions in SEB is not a completely direct measurement of heat and work, but it is in line with the majority of measurements of thermodynamics in small systems. We have presented in Sec.~\ref{Sec:Calorimetry+Bolometry} ways to measure heat bolometrically or calorimetrically, which are the most direct ways of determining heat experimentally. It is very rare, however, that heat can be measured directly: in this respect the recent experiment on MD in a single-electron circuit is an exception \cite{Koski2015}. In what follows we relate the transition statistics (in theory and experiment) to common fluctuation relations and to the analysis of Maxwell's Demons. The basic SEB is naturally the most simple single-electron circuit to consider, and more involved circuits are applied in experiments. The relevant expressions will be given along the way, as such circuits are presented.

\subsection{Fluctuation relations}
Fluctuation relations governing small systems even far from equilibrium conditions, are currently under active investigation \cite{bochkov1977general,bochkov1981nonlinear,Evans1993,Gallavotti1995,Jarzynski1997,Crooks1999,Seifert2005,Seifert2012_RepProgPhys}.
This is because various nano- and micro-scale systems have become experimentally feasible over the past years. Although fluctuation relations often serve just as sanity checks of the experiment, they also provide means to extract parameters of the studied system and, as advertised since early experiments, the equilibrium free energy of the system \cite{Collin2005,Alemany2011,Alemany2015}. The recent studies of MDs and information powered refrigerators yield interesting connections of information and energy via the newly found generalized fluctuation relations \cite{Sagawa2008,Sagawa2010}.

The general fluctuation relation (see \cite{Crooks1999,Seifert2005,Seifert2012_RepProgPhys} and references therein)
\begin{equation} \label{e8_Crooks}
P(\Delta S)/P_{R}(-\Delta S) = e^{\Delta S/k_B}
\end{equation}
relates the probability distribution $P(\Delta S)$ of stochastic entropy production $\Delta S$
in the forward process with the one $P_R(-\Delta S)$ in the reversed process
\footnote{There are some other variants for the same probability distribution $P(\Delta S)$ either in steady state conditions \cite{Evans1993, Gallavotti1995,Lebowitz1999} or in periodically-driven systems \cite{Shargel2009}.}.
This expression leads simply to the well known expectation value of the exponent of the entropy production in repeated experiments with a given protocol as
\begin{equation} \label{e9_JE}
\langle e^{-\Delta S/k_B}\rangle =1.
\end{equation}
These expressions have their counterparts for driven systems either initialized in equilibrium with the temperature $T$, where $\Delta S$ is replaced by $W_d/T$ \cite{Jarzynski1997},
or under steady state drive conditions, where in Eq. \eqref{e8_Crooks} the probabilities in numerator and denominator correspond to the same probability distribution \cite{Gallavotti1995,Lebowitz1999,Shargel2009,Schuler2005,Tietz2006}
\footnote{Some relations between generally- and periodically-driven or steady-state regimes are currently under investigations, see, e.g., \cite{Mandaiya2018}.}.
%Both these expressions have their counterparts for driven systems, where $\Delta S$ is replaced by $W_d$ and in Eq. \eqref{e8_Crooks} the probabilities in numerator and denominator correspond to opposite protocols in time of the driving field(s).

An alternative way to quantify the events of negative entropy production is to address its negative record statistics.
For stationary processes there have been several universal bounds derived for the statistics of the events of negative entropy production \cite{Chetrite2011,Neri2017,Pigolotti2017}.
However charge counting in single-island devices, such as SEB or SET, is not applicable to address the above mentioned observables.
Indeed, these devices, being a perfect toolbox for quantitative experimental
studies of stochastic thermodynamics~\cite{Pekola2015NatPhys},
do not provide information on the direction of electron tunneling, a feature that is key to characterize entropy production in the system in nonequilibrium steady state in contact with several reservoirs.
To access the negative record statistics one needs single-electron devices with {\it multiple} islands
allowing to measure the charge tunneling direction~\cite{Fujisawa2006,Kung2012,Singh2017} and thus provide a test bench for results of steady-state thermodynamics.
In this review we will focus mostly on single-island devices that are the simplest ones allowing to realize Maxwell's demons.

\section{%Classical Maxwell's Demons in single-electron circuits
CLASSICAL MAXWELL'S DEMONS IN SINGLE-ELECTRON CIRCUITS
}\label{Sec:MD}
\subsection{Introduction}

%\begin{itemize}
%  \item MD as a thought experiment
%  \item Szilard engine as a quantitative proposal + analogy to any 2-level system
%\end{itemize}

Fluctuation relations \eqref{e8_Crooks}, \eqref{e9_JE} considered above hint that the second law of thermodynamics is satisfied only on average and should be ``violated'' in a certain amount of realizations due to thermal fluctuations of the bath(s) coupled to the system.
This fact opens a way to utilize these thermal fluctuations challenging the second law by a thought experiment by Maxwell:
A human being, called Maxwell's demon, by measuring and controlling a microstate of the system can extract some work of it by decreasing a system stochastic entropy $S_{sys}$ without direct energy costs.
This paradoxical example has initiated long-standing debates of the concept of Maxwell's Demon and of the role of the information in statistical physics and feedback control (see,e.g., \cite{Leff2002,Maruyama2009,Parrondo2015} and references therein).

\begin{figure}
\includegraphics[width=0.6\textwidth]{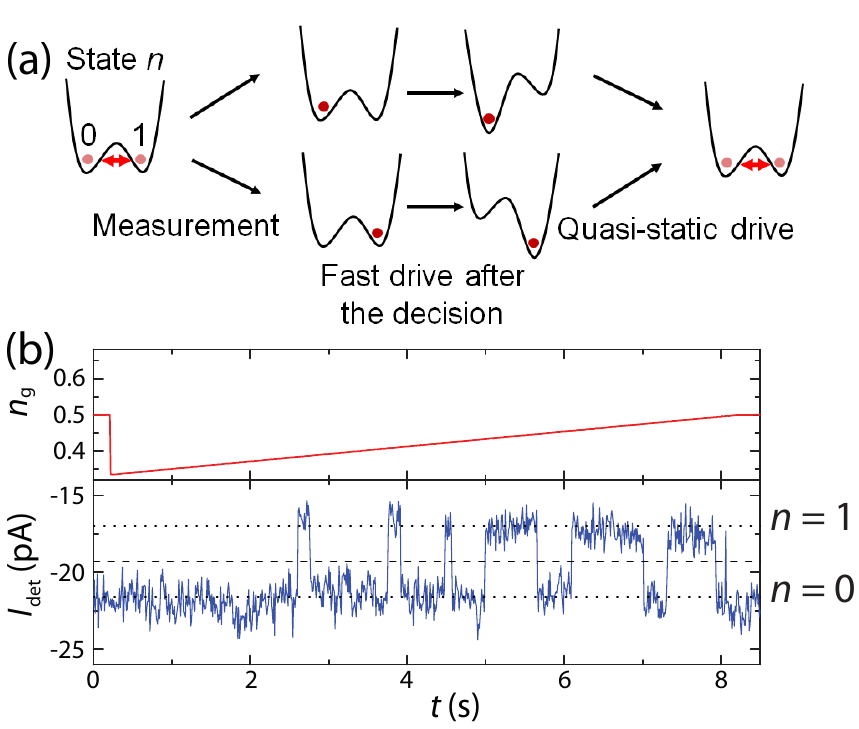}
\caption{(a)~The original proposal of Szilard engine in terms of a classical two-level system (shown as a particle in a two-well potential).
(b)~The top trace shows the applied gate voltage signal $n_g = C_g V_g/e$ providing a feedback.
The bottom time trace is the measured excess electron location signaled by the SET current $I_{det}$.
The corresponding values of $n$ are shown on the right.
Panel~(b) is adapted from \cite{Koski2014_PRL}.
}
\label{Fig:SE_a_Scheme_b_Protocol_c_Traj}
\end{figure}

\begin{marginnote}[120pt]
\entry{TLS}{two-level system}
\end{marginnote}

Leo Szilard has given a more quantitative example of information-mediated engine which relates a bit of information gathered by an observer with the maximal amount of work possible to extract from a system.
The main idea of this Szilard engine can be explained on an example of a classical two-level system (TLS) coupled to a single bath, see Fig.~\ref{Fig:SE_a_Scheme_b_Protocol_c_Traj}(a).
Consider the TLS initialized in the state with maximal average entropy $\langle S_{sys} \rangle = k_B \ln 2$ corresponding to equal probabilities of both states
\rev{$P_n = 1/2$, $n=0,1$.
%Due to bath thermal fluctuations the actual state of the system $n$ is a stochastic variable with the above mentioned realization probabilities.
The actual state of the system $n$ is a stochastic variable governed by bath thermal fluctuations.
After an ideal measurement of the state $n$, one can freeze TLS in it, e.g., %by squeezing the system phase space or
by tilting the potential
such that it makes %jumps to
the excited state %to be
nearly
%impossible
unreachable, $P_{n} = 1$, $P_{1-n} = 0$, see upper (lower) leg of Fig.~\ref{Fig:SE_a_Scheme_b_Protocol_c_Traj}(a) for $n=0(1)$.
}
In this process the detector stores one bit of information as $n = 0$ or $1$.
The averaged system entropy after the measurement becomes zero, $\langle S_{sys} \rangle = 0$.
To close the loop one can adiabatically return the system to the energy level crossing point with the equilibrium distribution $P_0 = P_1 = 1/2$ %($\langle S_{sys} \rangle = k_B \ln 2$)
and extract on average the work $\langle W \rangle = k_B T\langle\Delta S_{sys}\rangle = T \ln 2$.
As the TLS and the bath constitute a thermally isolated system $\Delta S\geq 0$
the process behind Szilard engine is nothing else, but a conversion of heat $-\langle Q \rangle \leq k_B T \ln 2$ extracted on average from a single bath into work. This process seems to be a realization of a perpetuum mobile of the second kind.

%\subsection{Information in thermodynamics}
%\begin{itemize}
%  \item Landauer's principle
%  \item Mutual information
%  \item Sagawa-Ueda relations $\langle e^{-\beta(W-\Delta F)-I}\rangle=1$ and $\langle e^{-\beta(W-\Delta F)}\rangle=\gamma<2$
%\end{itemize}

To resolve this obvious paradox one should take into account the observer's thermodynamics,
the amount of the information stored in its memory
and an unavoidable need to erase this information.
As imposed by Landauer's principle \cite{Landauer1961,Landauer1988} the energy dissipation needed to erase one bit of information is bounded by
$\langle Q_{det} \rangle \geq k_B T \ln 2$. This principle restores, in particular, the second law  for the Szilard engine
\begin{gather}
\langle Q_{det} \rangle + \langle Q \rangle \geq 0 \ .
\end{gather}

In more rigorous analysis one should take into account errors in the measurement: in particular, the outcome of the measurement $m$ being a stochastic variable is correlated with the actual state of the system $n$, but it does not necessarily coincide with it.
To quantify this correlation in the information theory the
\rev{%following quantity called
}
mutual information is introduced
\begin{gather}\label{e11_mutual_info}
I_M (n| m) = \ln P_{n, m} - \ln P_{n} - \ln P_{m} \ ,
\end{gather}
where $P_{n,m}$, $P_{n}$, $P_{m}$ are joint and marginal distributions of the actual state $n$ and of the measurement outcome $m$
\footnote{The detailed analysis of errors, optimization of the feedback, and the higher order tunneling processes in this system have been investigated, e.g., in \cite{Bergli2013,Sordal2017,Walldorf2017,Averin2017}.}.
The average mutual information $\langle I_M\rangle = \sum_{m,n }P_{n,m}I_M (n| m)$ is non-negative and has the upper bound of the Shannon's entropy $\langle S_{sys}/k_B \rangle$ obtained in the completely correlated case $m = n$.
$I_M (n| m)$ plays an important role in the information theory and, in particular, appears in the generalization of Jarzynski equality \eqref{e9_JE} for a feedback-controlled system as follows \cite{Sagawa2010}
\begin{gather}\label{e12_JE_Sagawa_Ueda}
\langle e^{-\beta(W-\Delta F)-I_M}\rangle=1 \ .
\end{gather}
Note that the standard Jarzynski equality is violated in this case $\langle e^{-\beta(W-\Delta F)}\rangle=\gamma_{JE}$ with $0<\gamma_{JE}<2$.
Consequently, the mutual information also places a lower (negative) bound on the average work dissipated in the system under feedback control  \cite{Sagawa2008}
\begin{gather}\label{e13_Wd>-I_M}
\langle W - \Delta F\rangle \geq - k_B T \langle I_M \rangle \ .
\end{gather}

\subsection{Szilard's Engine}
%\begin{itemize}
%  \item SEB as a 2-level system + protocol + Landauer's principle
%  \item Work distribution $P(W)$ with two peaks + errors
%  \item $\langle W\rangle \simeq -0.75 k_B T \ln 2$
%  \item Detector bandwidth (or filtering) as error level
%  \item Verification of Sagawa-Ueda relations $\langle e^{-\beta(W-\Delta F)-I}\rangle=1$ and $\langle e^{-\beta(W-\Delta F)}\rangle=\gamma<2$
%\end{itemize}

In last decade several realizations of Maxwell's Demons have been demonstrated in various systems ranging from colloidal particles \cite{Toyabe2010, Roldan2014} and photons \cite{Vidrighin2016} to complicated objects such as DNA molecules and organic polymers \cite{Ribezzi}.
Single-electronic devices being an easy realization of a classical few-level system
give an opportunity to realize the concept of information-mediated work extraction (see \cite{Koski2014_PRL,Koski2014_PNAS} for seminal experiments), which thereafter can be used for electronic cooling \cite{Koski2015,Chida2015} and squeezing of shot noise \cite{Wagner2016}.
Another advantage of single-electronic devices is their stability and robustness
%comparing to complicated biological system,
allowing one to repeat a drive protocol many times and collecting large statistics in one sample.
Overall, despite all simplicity of single-electronic devices, they can work as a benchmark for testing various concepts of stochastic thermodynamics \cite{Saira2012,Koski2013}, counting statistics \cite{Singh2016}, and even simulate multifractality of the critical wave functions in the vicinity of the Anderson transition \cite{Khaymovich2015}.

The first MD in single-electronics realized in Refs.~\cite{Koski2014_PRL,Koski2014_PNAS} uses a SEB as a feedback-controlled system
with the very same protocol of Szilard engine as the one discussed above and shown in Fig.~\ref{Fig:SE_a_Scheme_b_Protocol_c_Traj}(a).
The initial state of the maximal average system entropy is realized at the control parameter $n_g=1/2$ where the energies $U(0, n_g)$ and $U(1,n_g)$,
Eq.~\eqref{e3_U(n,ng)}, are degenerate.
The measurement of the microscopical state $n$ performed by a capacitively coupled single-electron transistor (SET) working as a charge detector,
see Fig.~\ref{Fig:Calorimetry+NIS+SEB}(d), gives the outcome $m$.
The feedback freezing the SEB in its ground state is applied by rapid driving of $n_g$ towards $m$ value.
Eventually $n_g$ is slowly returned to the degeneracy point, see several realizations of this protocol in Fig.~\ref{Fig:SE_a_Scheme_b_Protocol_c_Traj}(b).

\begin{figure}
\includegraphics[width=0.8\textwidth]{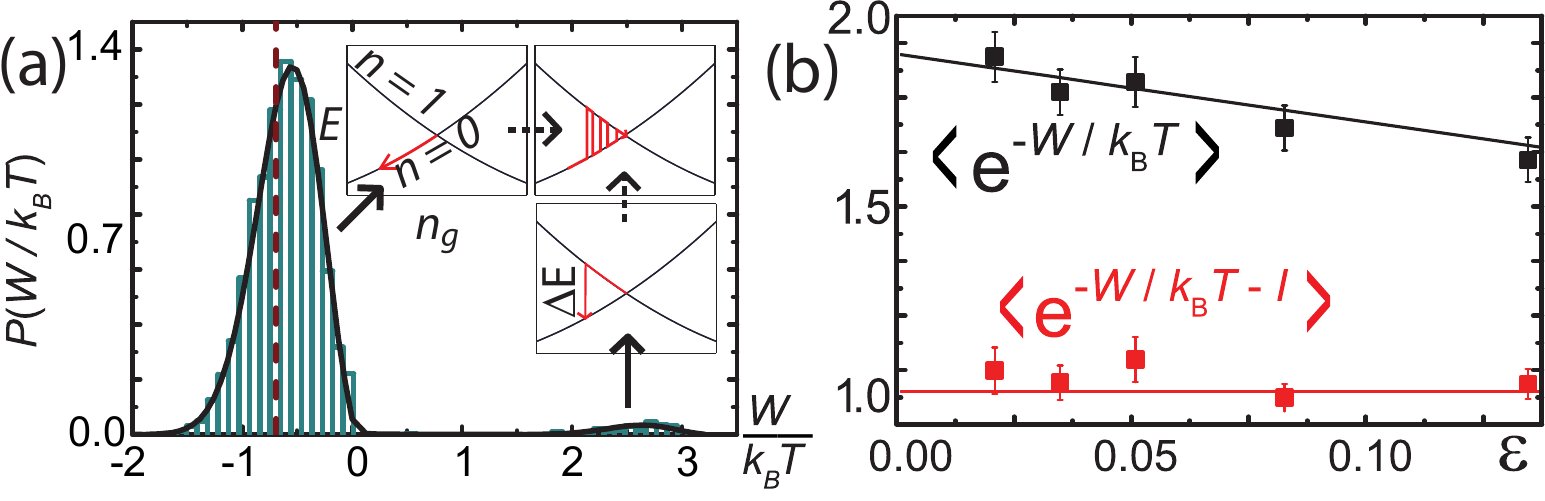}
\caption{(a)~Distribution $P(W)$ of experimentally observed work under feedback control is shown as a histogram.
The numerical results (black solid line) are in good agreement with the experimental data.
Insets show sketch of two processes corresponding to two peaks in $P(W)$.
(b)~Test of the standard Jarzynski equality \eqref{e9_JE} and generalized one \eqref{e12_JE_Sagawa_Ueda} with mutual information.
The error bars include the statistical error, as well as the uncertainty in the measured value of $E_C/ k_B T$.
Figure is adapted from \cite{Koski2014_PRL,Koski2014_PNAS}.
}
\label{Fig:SE_a_P(W)_b_Sagawa_Ueda}
\end{figure}

To determine the work \eqref{e5_1st_law} equal to the heat \eqref{e4_Q_sum_over_jumps} in each realization of this cyclic process
the SET is constantly monitoring the charge state $n$ reconstructing charge trajectories $n(t)$.
Repeating this process many times, one can extract the distribution $P(W)$ of applied work, Fig.~\ref{Fig:SE_a_P(W)_b_Sagawa_Ueda}(a),
which concentrates mostly on the negative side close to the ideal value $-k_B T \ln 2$
(the top two insets show the corresponding process).
%Because the slow part of the drive is not fully adiabatic,
Due to not fully adiabatic slow part of the drive,
cycle-to-cycle fluctuations of work form continuous $P(W)$.
A small bump far on the positive side of the distribution is related to errors in measurement, $m\ne n$, and/or delay in feedback control,
which send the system to the excited state with excess charging energy dissipated eventually to the bath
(see two right insets of Fig.~\ref{Fig:SE_a_P(W)_b_Sagawa_Ueda}(a)).
The optimization of the control parameter value $n_g$ reached in the fast part of the drive gives
the overall average work $\langle W\rangle = -0.75 k_B T \ln 2$ quite close to its ideal value, meaning that up to $75$~\% of the collected information can be extracted as work from the bath.

To uncover the role of the mutual information \eqref{e11_mutual_info} in this experiment the error probability
$P(m\ne n)\equiv\epsilon\leq 1/2$ is efficiently tuned by changing of the SET detector bandwidth.
Finite $\epsilon$ determines the mutual information as follows $I_M(n=m) = \ln[2(1-\epsilon)]\geq 0$ and $I_M(n\ne m) = \ln (2\epsilon)\leq 0$.
Because of the applied feedback the standard Jarzynski equality is violated $\langle e^{-\beta(W-\Delta F)}\rangle=\gamma_{JE}$
with the parameter $0<\gamma_{JE}<2$ decreasing with the error probability $\epsilon$, while its generalized version \eqref{e12_JE_Sagawa_Ueda}
including the mutual information is still satisfied with experimental accuracy of $8$~\%,
Fig.~\ref{Fig:SE_a_P(W)_b_Sagawa_Ueda}(b).
The average mutual information $\langle I_M \rangle = -\epsilon \ln \epsilon - (1-\epsilon)\ln(1-\epsilon)$ also places the upper bound
\eqref{e13_Wd>-I_M} on the average extracted $W-\Delta F$ showing that the feedback efficiency is decreasing as errors grows.%, Fig.~\ref{Fig:MD_Wd>-I_M}.

\subsection{Autonomous single-electron Maxwell's Demon}
%\begin{itemize}
%  \item Main idea on the level of electron trapping in SET
%  \item 2 coupled SETs as a realization-1:  Simple description of charging energy (see ppt Oslo)
%  \item 2 coupled SETs as a realization-2: Coulomb interaction + different resistances as a principle + simple analysis of $I$ and $\dot{Q}_S$ in terms of $P_{odd}$ and $P_{even}$
%  \item Different regimes: SET-cooler and an autonomous MD
%  \item Demons' thermodynamics: $\dot{Q}_d \geq k_B T I_{M,d}$.
%\end{itemize}

In all previous examples, feedback-controlled systems were in the focus of consideration,
while the role of the Demon itself was played by an external agent, whose thermodynamics related to the information gathering and
\rev{%the memory
}
erasure was implicitly assumed to be valid.
Some aspects of MD thermodynamics
have been observed in various mesoscopic systems on the level of verification of Landauer's principle
\cite{Berut2012, Orlov2012, Jun2014, Hong2016, Gavrilov2017}, but it has been done separately from the actual Demon's operation.
All fluctuation relations for work and entropy production, see, e.g., Eqs.~\eqref{e8_Crooks}, \eqref{e9_JE}, \eqref{e12_JE_Sagawa_Ueda}, describe only the system keeping the Demon's dynamics beyond the scope.
Another issue with most above mentioned MD experiments is that the heat extracted from the bath is measured indirectly (due to fast relaxation rates) and not transformed into some useful work such as increasing of the free energy or charging of a battery.

To access both the Demon's thermodynamics and the effect of the extracted work it is natural to implement a MD to the same configuration as a feedback-controlled System and design it in such a way that the information processing and feedback are carried out autonomously \cite{Strasberg2013,Horowitz2014,Shiraishi2015}.
On one hand, this
\rev{%gives an opportunity
allows one to observe %directly
}the heat dissipation related to the memory erasure in the standalone MD and compare it with the mutual information gained.
On the other hand, autonomous character of the Demon's operation allows to increase the %heat and information flows between
operating frequencies of the System
\rev{%, the bath,
}and the Demon and direct the heat extracted from the bath to some useful work with a noticeable power, which can be
\rev{%also
}directly measured.
Note that here we restrict our consideration to autonomous MDs without a separate memory register to access directly the heat dissipated due to information erasure.

\begin{figure}
\includegraphics[width=0.6\textwidth]{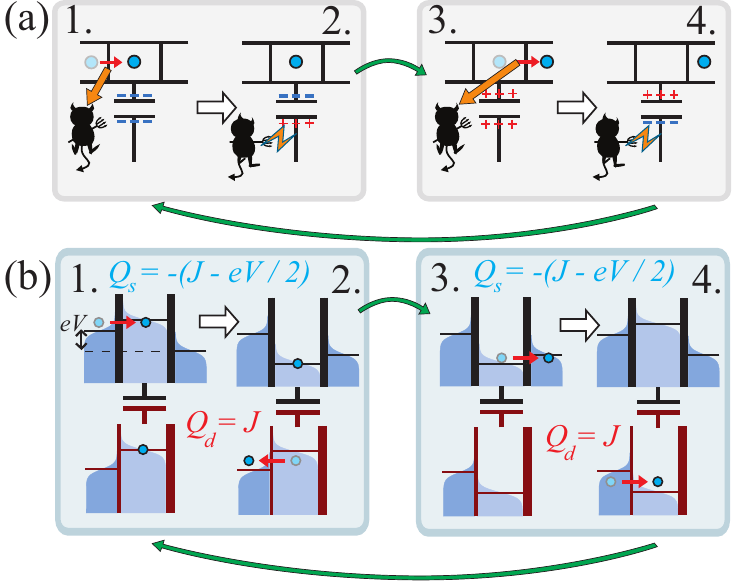}
\caption{
\rev{
%Operation principle of an autonomous Maxwell's demon.
%(a)~The Demon monitors a System SET for electrons that tunnel into (out of) the island in panel 1(3).
%Then it applies a feedback by putting a positive (negative) potential to trap (repel) the electrons, panel 2(4).
%%Coulomb blockade ensures that only either one or zero electrons reside in the System island.
%The electrons are always tunneling against the potential induced by the Demon, and therefore the System cools down.
(a) Operation principle of an autonomous Maxwell's demon.
Due to the potential applied by the Demon
the electrons in the system are always tunneling against this potential, and therefore the System cools down.
}
(b)~Energy diagram of autonomous Maxwell's demon setup
built of a System SET (shown in black) under voltage bias $V$ coupled to a Demon's SEB (shown in red).
\rev{%The conduction electrons of the System follow Fermi distribution (see blue areas for the corresponding occupation numbers), providing
Finite electronic temperature of the System provides}
electrons that can overcome the energy cost $J - eV/2$, cooling down the System at the same time.
The energy $J$ is dissipated by the Demon as it reacts (panels 1 and 3), changing the projected energy cost experienced by the electron tunneling
in the System from $-J - eV/2$ to $J - eV/2$.
Panel~(a) is adapted from \cite{Koski2015}.
}
\label{Fig:AutoMD_a_Protocol_b_Energetics_SET+SEB}
\end{figure}

In the seminal experiment on an autonomous MD \cite{Koski2015} a biased SET plays a role of a feedback-controlled system where an electron can tunnel between the Coulomb blockaded island and two leads changing the charge state between $n=0$ and $1$ and cooling or heating the corresponding junction depending on the difference in effective chemical potentials in accordance with the Joule's law.
The main idea of the autonomous Maxwell's Demon behind this is the following, Fig.~\ref{Fig:AutoMD_a_Protocol_b_Energetics_SET+SEB}(a).
As soon as something is happening in the System to be observed there is a faster detector which reacts immediately
and gives the feedback to the monitored System. %erasing at the same time its own memory.
%So that it indeed creates a cycle where
%As in the System an electron tunnels uphill into the island the Demon immediately moves down the level on the System SET island by applying an attractive electrostatic potential and traps the electron there.
When an electron tunnels in the System uphill into the island, the Demon reacts immediately.
Due to a tunneling event in it, it moves down the level on the System SET island via an attractive
electrostatic potential trapping the electron there.
In this case to escape from the System island the electron again has to tunnel uphill and the Demon closes the loop by applying a repulsive potential on the island
moving the island level up again.
Overall the System cools down whereas the detector should in fact heat up due to continuous memory erasure.

In the actual experiment the Demon is made by an unbiased SET (equivalent to SEB) with $N$ excess electrons on the island,
capacitively coupled to the System SET \footnote{The configuration resembles theoretical proposals of heat-to-current convertors on quantum dots \cite{Sanchez2011,Sanchez2012} realized in several experiments up to now \cite{Thierschmann2015NatNano,Thierschmann2015NJP}.}. The resistance of the detector junctions $R_d$ determining its operating frequency by tunneling rates
is made much smaller than the one in the System $R_s$, $R_d\ll R_s$.\footnote{Here for simplicity we consider both detector (Demon) and System with equal left and right tunnel junctions.}
This allows MD to measure and apply feedback faster than the System evolves.
The energy of the supersystem of two coupled single-electronic devices
\begin{equation}\label{e14_U(n,N,n_g,N_g)}
U(n,N,n_g,N_g)=E_{C,s}(n-n_g)^2 + E_{C,d}(N-N_g)^2 + 2J(n-n_g)(N-N_g).
\end{equation}
contains electrostatic energies \eqref{e3_U(n,ng)} of the System itself and the Demon as the first two terms, with
the corresponding polarization charges $0<n_g,N_g<1$ in units of $e$. \rev{%induced by the constant gate voltages.
}
The latter term describes the mutual Coulomb interaction between subsystems.
The expressions for the corresponding Coulomb energies $E_{C,s}$, $E_{C,d}$, and $J$ are given in \cite{SM}.
%The corresponding Coulomb energies
%$E_{C,s} = e^2 C_{\Sigma,d}/2(C_{\Sigma,s}C_{\Sigma,d}-C_m^2)$, $E_{C,d}=e^2 C_{\Sigma,s}/2(C_{\Sigma,s}C_{\Sigma,d}-C_m^2)$, and
%$J=e^2 C_{m}/2(C_{\Sigma,s}C_{\Sigma,d}-C_m^2)$ depend on the total capacitances $C_{\Sigma,s,d}$ of the islands and on their mutual capacitance $C_m$.

As the Demon is assumed to be autonomous, the supersystem is put in steady state conditions
by keeping both control parameters $n_g$ and $N_g$ to be constant.
A rather small constant bias voltage $eV\ll E_{C,s},E_{C,d}$ applied to the System makes the overall heat dissipation rate in the Demon
$\dot Q_d$ and in the System $\dot Q_s$ satisfy Joule's law $\dot Q_d+\dot Q_s = I V$, where $I$ is the steady state System current.
To measure the work extracted by Demon, one can direct this work, e.g., to the cooling of the System which is in contact just with one bath.
The resulting steady state heat flows in both subsystems $\dot Q_d$, $\dot Q_s$ %arisen from continuous Demon's operation
can be directly measured via changes in the temperatures of System $T_L$, $T_R$ and detector $T_{det}$ electrodes.\footnote{\rev{For thermal isolation of the System and Demon from the remaining world, superconducting leads were used in the experiment.}}

The values of control parameters $n_g$ and $N_g$ can significantly change the mutual dynamics of the System and MD.
%As usual in Coulomb blockaded devices,
Polarization charge $n_g$ of the biased System SET controls the current $I$ flowing through the System, suppressing it at integer values and
putting it to the maximum at half-integers.
Unlike this, in the unbiased Demon SET, the parameter $N_g$ governs mostly the coupling of the System and MD tuning the amplitude
$2J\min(N_g,1-N_g)$ of the mutual electrostatic energy variation as $n$ changes between $0$ and $1$ and $N$ is in the corresponding ground state.
\rev{
%Thus, integer $N_g$ values switch off the applied feedback, while $N_g = 1/2$ maximizes it.
}

\begin{figure}
\includegraphics[width=0.6\textwidth]{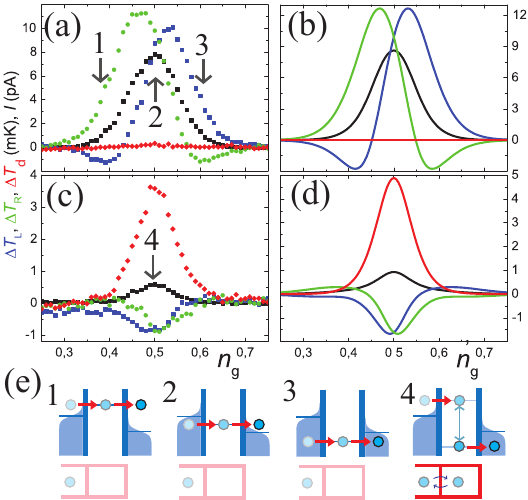}
\caption{
Maxwell's demon operation with the bias voltage $V = 20$~$\mu$V.
(a, b)~the measurement of System current $I$, System $T_{L,R}$ and Demon's $T_{det}$ temperatures at operation point
$N_g = 0$, where the Demon is not interacting with the System.
Measured data is shown on the left, and simulated data on the right. Color code is given in the $y$-axis labels.
One-sided cooling can be observed in the System \cite{Pekola2014,Feshchenko2014}, however the total heat generated is positive.
(c, d)~In the feedback regime at $N_g = 0.5$, both $T_L$ and
$T_R$ drop below the base temperature value, showing that the System is thoroughly cooled
down by the feedback operation by the Demon. Heat is generated in the Demon in turn.
Panels~(e) show the energetics at the different operation points indicated by numbers in panels (a) and (c).
Figure adapted from \cite{Koski2015}.
}
\label{Fig:AutoMD_results_a_SET_cooler_b_Feedback_c_cases}
\end{figure}

\rev{%Consider first the feedback switched off, $N_g=0$. In this case the biased System SET is decoupled from the feedback.
In the case of $N_g=0$ the biased System SET is decoupled from the feedback.
}
The bias value determines the dominant tunneling direction of electrons.
The parameter $n_g$ governs the current flowing through it by changing the effective energy level of an electron in the SET island
(see panels 1-3 in Fig.~\ref{Fig:AutoMD_results_a_SET_cooler_b_Feedback_c_cases}(e)).
If this level is in between electrode biases, the electron tunneling in both junctions occurs with a positive energy gain leading to heat dissipation
in both electrodes.
Otherwise, if the level is slightly above the upper bias or slightly below the lower one,
an electron has to tunnel uphill in one of two junctions absorbing the energy from thermal fluctuations of the bath and,
thus, cooling the corresponding electrode down.\footnote{ \rev{Note that at the same time the current through the device is suppressed as thermally-activated tunneling is exponentially slower than that with the positive energy gain.}}
\rev{
%Nevertheless, the total heat dissipated in two sequential tunneling events at any $n_g$ should satisfy the Joule's law $\dot Q_s = I V > 0$, thus, the overall dissipation in the System is non-negative.
Nevertheless, due to the Joule's law $\dot Q_s = I V > 0$ the total heat dissipated in the System is non-negative at any $n_g$.
}
This effect of the cooling in the normal SET  has been
discovered in \cite{Pekola2014}, demonstrated in \cite{Feshchenko2014}
and reproduced in \cite{Koski2015} for $N_g=0$,
see Fig.~\ref{Fig:AutoMD_results_a_SET_cooler_b_Feedback_c_cases}(a, b).
\rev{
%The remarkable thing in it is that %either left or right electrode is heated up at any gate voltage position.
%there is no way to cool down both electrodes simultaneously.
}

However, one can
cool down both electrodes of the System
%\rev{make the overall dissipation in the System to be negative}
by switching on the feedback.
In this case the temperatures of the System electrodes have negative deviations while Demon heats up significantly,
see Fig.~\ref{Fig:AutoMD_results_a_SET_cooler_b_Feedback_c_cases}(c, d).
As each tunneling event in the System happens with heat absorbtion from the bath, the System current is suppressed compared to case without any feedback.
To get the qualitative idea of this process we consider the simplest case of maximal coupling between subsystems, $N_g = 1/2$,
and current flowing through the System, $n_g=1/2$
(see panel 4 in Fig.~\ref{Fig:AutoMD_results_a_SET_cooler_b_Feedback_c_cases}(e)).
We also assume temperatures to be quite small $k_B T_L, k_B T_R, k_B T_{det}\ll E_{C,s},E_{C,d}$ with respect to the charging energies.
In this case $n$ and $N$ are limited to two possible values $0$ and $1$
and the electrostatic energy \eqref{e14_U(n,N,n_g,N_g)} depends on these variables only via the last term
%give just a constant offset as $N,n = \overline{0,1}$ and the
\begin{gather}
\delta U (n,N) = U(n,N,1/2,1/2)-E_{C,s}/4 - E_{C,d}/4 = J(2n-1)(2N-1)/2 \ .
\end{gather}
This forms a doubly degenerate \rev{%charge neutral
ground state $(0,1)$, $(1,0)$, %with the energy $\delta U = -J/2$,
and %a doublet of charged
excited state $(0,0)$ and $(1,1)$, with the corresponding energies $\delta U = J/2, -J/2$ and steady-state probabilities
%Considering for simplicity the equal System junctions and temperatures $T_L=T_R\equiv T_s$,
%one can prove that steady state probabilities
$P_{n,N}$ }%to realize a state of each doublet are equal
\begin{subequations}
\begin{align}
P_{0,1} = P_{1,0}\equiv P_g/2 \ ,\\
P_{0,0} = P_{1,1}\equiv P_e/2 \ .
\end{align}
\end{subequations}
Here $P_g = \Gamma_+/(\Gamma_+ +\Gamma_-)$, $P_e = \Gamma_-/(\Gamma_+ +\Gamma_-)$,
and the upper (lower) sign in $\Gamma_\pm = \Gamma_s(\pm J_+)+\Gamma_s(\pm J_-)+\Gamma_d(\pm J)$
corresponds to the tunneling to the ground (excited) state.
The System (detector) tunneling rates $\Gamma_{s (d)}$ are given by Eq.~\eqref{e0_Gamma_rate} with normal metal DOSes $n_1 = n_2 = 1$ and $R_T$ substituted by $R_{s (d)}$.
The energy gains $J_\pm = J\pm eV/2$ in tunneling events in the System are subject to the bias voltages of the left $V/2$ and right $-V/2$ lead, respectively
(for complete energetics see Fig.~\ref{Fig:AutoMD_results_a_SET_cooler_b_Feedback_c_cases}(c, d)).
Due to a small detector resistance $R_d\ll R_s$ and a weak $\delta E$-dependence of the rates $\Gamma_{s(d)}(\delta E)$ at $\delta E>0$
the relaxation rate to the ground state is mostly dominated by the tunneling in the detector $\Gamma_+\simeq \Gamma_d(J)$.
At the same time it is mostly the tunneling in the System which brings the total supersystem to the excited state $\Gamma_- \simeq \Gamma_s(-J_-)\ll \Gamma_+$,
due to the exponential suppression of %the rates according to the local detailed balance
thermally-activated rates
$\Gamma_{s(d)}(-\delta E) = e^{-\delta E/k_B T_{s(d)}}\Gamma_{s(d)}(\delta E)$.
Ideally when all System tunneling events contribute only to the excitation rate $\Gamma_-$, while tunneling in the detector relaxes the total supersystem back,
the System itself should absorb the energy of $2J-eV$ from the bath and MD has to dissipate $2J$ per each electron transferred from left to right electrode.
Thus, the ideal heat generation rates in the System and in the Demon take the forms
$\dot Q_s^{ideal} = -(2J/e-V)I$ and $\dot Q_d^{ideal} = 2 J I/e$. %, respectively.
Here and further we consider $J-eV/2>0$.\rev{\footnote{The general (non-ideal) case is considered in details in \cite{SM,Koski2015}.}
Note that in any case the average System current is flowing towards bias
$2I/e = P_g \left[\Gamma_s(-J_-)-\Gamma_s(-J_+)\right]+P_e\left[\Gamma_s(J_+)-\Gamma_s(J_-)\right]>0$.}

Figure~\ref{Fig:AutoMD_results_a_SET_cooler_b_Feedback_c_cases}(c, d) shows that temperatures $T_L$, $T_R$ (in blue and green) of both System leads goes below the bath temperature value $T$ simultaneously,
while the System current (in black) is suppressed compared to the case of $N_g = 0$.
The best efficiency of order of 50\%, $\dot Q_d \simeq 0.5 \dot Q_d^{ideal}$ is experimentally achieved
near degeneracy point of the System, $n_g=1/2$, where the maximal current is flowing through the System.

\rev{
Although energetically this device follows Joule's law, it is the information flow between the System and MD that permits
the decrease of System entropy leading to its cooling.
According to \cite{Horowitz2014} the mutual information rates place an upper bound to
the System heat absorbtion rate $-\dot Q_s \leq k_B T_s \dot I_{M,d}$ and the lower bound to the heat rate generated in MD $\dot Q_d \geq k_B T_d \dot I_{M,d}$.
As the process is cyclic the average mutual information is conserved and the contribution to its rate governed by the System tunneling is $\dot I_{M,s}=-\dot I_{M,d}$.
The inequality $\dot Q_d \geq k_B T_d \dot I_{M,d}$ is saturated in the limit $R_d\ll R_s$ as in this case the probability distribution has a thermal form
$\ln(P_g/P_e) = 2J/k_B T_d$ (see \cite{SM} for details).
In the experiment the above mentioned inequality $\dot{Q}_d \geq k_B T_d I_{M,d}$ differs from equality only by $15$~\%,
which verifies the theoretical predictions for MD thermodynamics.

%To conclude, in this section we have demonstrated that the single-electron devices give a good benchmark for testing classical concepts of nonequilibrium
%statistical physics and thermodynamics, in particular, in the presence of feedback control.
%By easy combination of several single-electronic devices one can access not only thermodynamical properties of the system of interest,
%but also realize an autonomous feedback controlling device and observe its dynamics.
}

\section{%Thermodynamics in open quantum systems
THERMODYNAMICS IN OPEN QUANTUM SYSTEMS
}\label{Sec:ThermDyn_in_OQS}

\subsection{Stochastic thermodynamics of qubits}\label{Sec:Qubit_Th_dyn}

Stochastic thermodynamics of superconducting qubits is expected to be based on detection of single micro-wave photons, or in some schemes on measurement of the state of the qubit itself (see, e.g., \cite{Esposito2009,Campisi2011} and references therein). An important difference as compared to classical stochastic thermodynamics is that the measurement back-action has naturally a more fundamental meaning in quantum mechanics. In what follows we discuss the most obvious state measurement of the qubit and calorimetric detection of heat, the latter of which is most compatible with the spirit of traditional thermodynamics.

Quantum trajectories provide a basis to understand stochastic dynamics of qubits \cite{Dalibard1992}. On one hand they are a computational tool, but on the other they offer a way to represent the true dynamics in a single realization of an experiment. By averaging many such trajectories under identical experimental conditions and realizations, one obtains what is familiar from the standard master equation approach for the density matrix of an open quantum system. Thus this method based on quantum jumps looks like an ideal tool to treat problems in stochastic quantum processes; moreover, associating heat to jumps equivalent to the energy released in a transition process ($\pm$ the energy splitting of the qubit), it appears as a means to master the energy exchanges in quantum thermodynamic problems. This approach was harnessed in Refs. \cite{Hekking2013} and \cite{Horowitz2013}. In \cite{Hekking2013}, particular attention was given to a qubit driven at its resonance frequency over a few periods in each realization. The Jarzynski fluctuation relation was found to be fulfilled in general in the limit of coupling the qubit weakly to its equilibrium environment (a bath at a given temperature), bringing confidence to the physical correctness of the method. Further theoretical studies confirm this conclusion under different driving protocols \cite{Suomela2016,Pekola2016JLTP,Pekola2015PRE,Kupiainen2016}.

\subsection{Potential detection of single microwave photons by nano-calorimetry}\label{Sec:Calorimetry_single_photon_detection}
A way to detect quantum trajectories for thermodynamics experiments in a laboratory would be a calorimetry measurement of a mesoscopic bath \cite{Pekola2013}. A possible scenario is to couple a small metallic or semiconducting absorber as a termination of a coplanar resonator of a superconducting (transmon) qubit, as realized in several experiments up to now \cite{Partanen2017,Ronzani2018}. Yet, in order to be able to detect the quantum jumps, one needs to fulfil two conditions not realized in the earlier qubit set-ups in this context: (i) One needs to measure the temperature of the absorber fast \cite{Gasparinetti2015,Zgirski2017,Wang2018}, with a bandwidth exceeding the thermal relaxation rate of the electrons in the absorber, typically the electron-phonon rate which is of the order of tens of micro-seconds at low ($\lesssim 100$ mK) temperatures. (ii) The measurement is to be made at low temperature to decrease thermal noise and the heat capacity of the absorber. (iii) The absorber needs to be made physically small %though technically feasible,
in order to secure small enough heat capacity $C_e$ for sufficient thermal signal upon photon absorption event.

To be more quantitative, the energy resolution $\delta \epsilon$ of a calorimeter is given by $\delta \epsilon =\sqrt{G_{\rm th}C_e S_T}$, where $G_{\rm th}$ is the thermal conductance of the absorber to the bath, and $S_T$ is the spectral density of temperature noise at low frequencies, determined either by the measurement set-up or by natural temperature fluctuations. In the latter case, which presents the fundamental lower bound when the detector is ideal, the energy resolution is given by $\delta \epsilon = \sqrt{k_B C_e}T$ at temperature $T$. For realistic parameters, we may take the free-electron heat capacity of a metallic (e.g. copper) absorber of volume $10^{-21}$ m$^3$, at a temperature of 10 mK, which yields $\delta \epsilon /k_B \lesssim 0.1 $ K, which is few times smaller than the energy of the photons emitted/absorbed by a typical superconducting qubit. Therefore calorimetric single microwave photon detection looks feasible although challenging with a modest ($2 ... 10$) signal-to-noise ratio. Further improvements can be expected by using absorbers with lower electron density (semiconductors, two-dimensional conductors), i.e. with lower heat capacity.

\subsection{Qubit-based quantum heat engines and refrigerators}
A quantum heat engine is a quantum-mechanical device converting the heat flow between hot and cold reservoirs to useful work (see, e.g., \cite{Kosloff2014} for review).
The refrigerator is a kind of an inverted heat engine transferring heat from cold to hot reservoir by means of work performed on it.
Recently several theoretical proposals of quantum heat engines \cite{Alicki1979,Campisi2016,Hofer2016_PRB93,Kosloff2014,Scully2003,Quan2007,Marchegiani2016,Uzdin2015,Campisi2017}
and refrigerators \cite{Abah2016,Brandner2016,Niskanen2007,Hofer2016_PRB94,Karimi2016,Karimi2017} have been suggested and a few experimental realizations are given \cite{Rossnagel2016}.
Like for the classical heat engines and refrigerators the efficiency of their quantum counterparts is limited by the Carnot efficiency~\cite{Alicki1979,Campisi2014}.
However, such intrinsically quantum phenomena as entanglement add some specifics to the consideration of quantum systems.
Although typically the quantum nature of heat engines diminishes their efficiency \cite{Brandner2016,Karimi2016}, on the other hand it may squeeze the fluctuations of charge \cite{Wagner2016} and heat \cite{Ptaszynski} properties.

\begin{figure}
\includegraphics[width=0.8\textwidth]{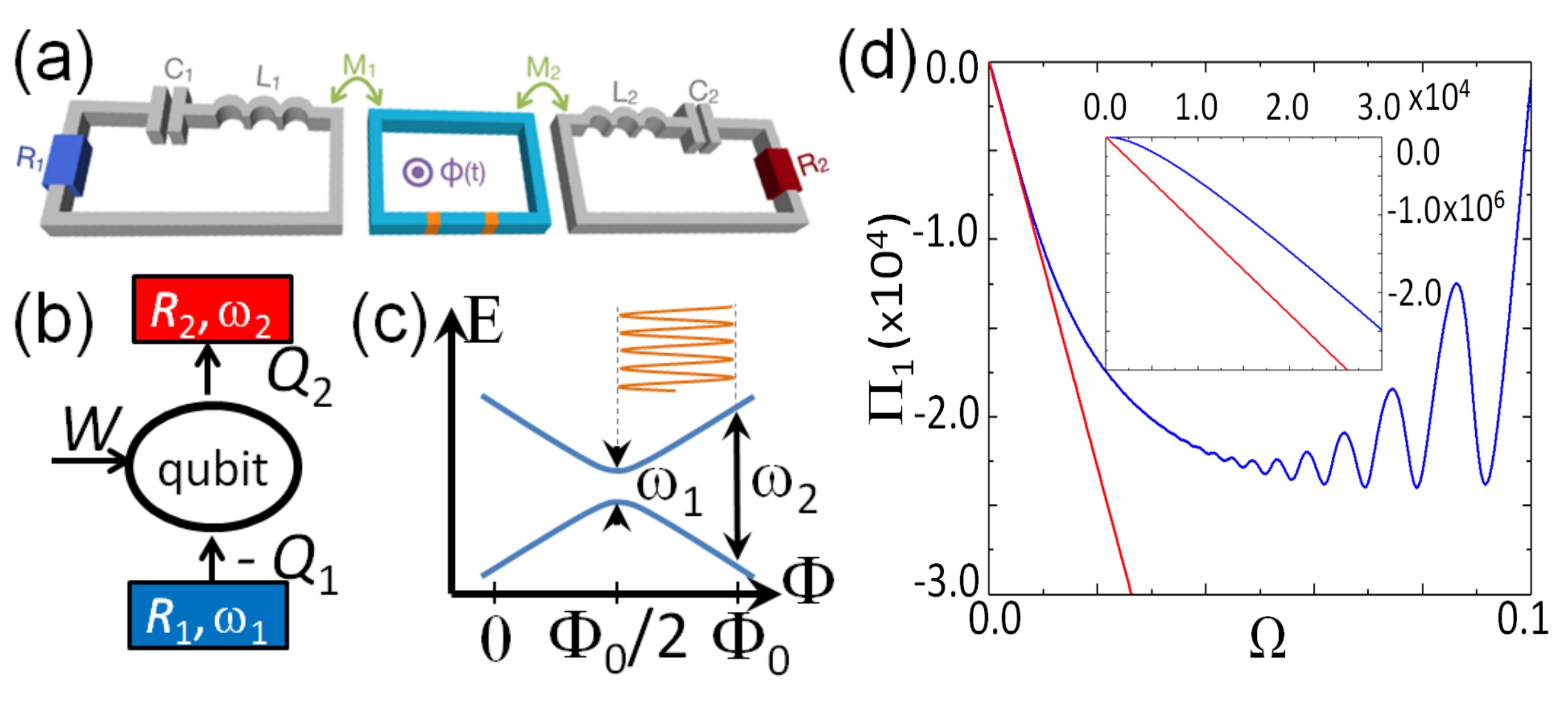}
\caption{
(a)~Schematics of a quantum refrigerator. The cold ($R_1$) and hot ($R_2$) resistors are shown by blue and red colors, respectively;
(b)~Heat and work flow schematics of the system. The resistors $R_{k}$ and frequencies $\omega_{k} \simeq (L_k C_k)^{-1/2}$, $k=1,2$, correspond to the circuit elements in panel (a);
(c)~Qubit energy diagram with the flux drive $\Phi(t)$ performing the work $W$ to the system.
(d)~The normalized cooling power $\Pi_1 = \hbar P_1/E_0^2$ of the cold reservoir $R_1$ versus dimensionless driving frequency $\Omega = 2\pi\hbar f/E_0$ showing different regimes of the quantum Otto refrigerator (see the text).
The inset shows the zoomed small frequency limit corresponding to the quadratic in $\Omega$ cooling power.
The parameters are the couplings $g = 4 E_0^2 M_i^2/(\hbar \Phi_0^2 R_i) = 1$, the quality factors $Q = \sqrt{L_i/C_i}/R_i = 30$, the temperatures $k_B T_i/E_0= 0.3$, with $i=1,2$, and the minimal qubit splitting $\hbar \omega_1/2E_0 = 0.15$.
Here $\Phi_0 = h/2e$ is the superconducting flux quantum and $E_0$ is the overall energy scale of the qubit.
This figure is based on the analysis and numerics by \cite{Karimi2016}.
}
\label{Fig:Otto_engine}
\end{figure}

In this section we address quantum heat engines and refrigerators based on qubits.
An early work on the qubit-based quantum heat engine \cite{Niskanen2007} considers a cyclic quantum Otto refrigerator consisting of
a flux qubit inductively coupled to two low- ($\omega_1$) and high- ($\omega_2$) frequency $LC$-resonators with resistors $R_{1,2}$ as heat baths placed at high ($T_1$) and low ($T_2$) temperatures, respectively, Fig.~\ref{Fig:Otto_engine}(a).
The mutual inductances are $M_{1,2}$, respectively.
The small damping $R_k\ll (L_k/C_k)^{1/2}$ of $LC$-resonators together with a strong coupling brings the selectivity in qubit relaxation, namely,
the qubit is coupled to $k$th resonator only in the vicinity of the corresponding frequency $\omega_k \simeq (L_k C_k)^{-1/2}$.
The adiabatic drive realized as a time-dependent flux $\Phi(t)$ with a period $1/f$ tunes the qubit energy splitting $\Delta E$ between the resonator frequencies $\omega_k$ (Fig.~\ref{Fig:Otto_engine}(c)) and performs work on the system (Fig.~\ref{Fig:Otto_engine}(b)).
As the qubit relaxation in the resonance is quite strong the population of qubit levels at $\Delta E = \omega_k$ follows the Gibbs distribution with the corresponding $k$th resistor temperature $T_k$.
In this approximation the refrigerator works provided the temperature of the cold resistor $T_1>\omega_1 T_2/\omega_2$ is not too small.

The paper \cite{Karimi2016} is devoted to the detailed analysis of operating the Otto refrigerator in different frequency regimes.
There using the standard quantum master equation written in the instantaneous eigenbasis \cite{Breuer2002,Pekola2016PRB} and ignoring pure dephasing due to intentionally large relaxation rates close to $\Delta E =\omega_k$, the authors have found three distinct regimes:
(i) (nearly) adiabatic regime,
(ii) ideal Otto cycle, and
(iii) high-frequency regime.
The (nearly) adiabatic regime shows the quadratic deviation of the powers dissipated to the hot and cold reservoirs with $f$ and the negative influence of quantum coherence on the engine performance for any drive profile $\Phi(t)$ (see the inset of Fig.~\ref{Fig:Otto_engine}(d)).
In the ideal Otto cycle regime, when the qubit thermalizes both at $\Delta E = \omega_1$ and $\omega_2$ and its population is kept intact between resonances $\omega_1<\Delta E<\omega_2$, the cooling power is approximately linear in $f$ in agreement with \cite{Niskanen2007}.
The hallmark of the high-frequency limit is the coherent qubit oscillations present in the $f$-dependence of the powers dissipated in the two baths (see Fig.~\ref{Fig:Otto_engine}).
This regime spreads in the range $\omega_1\lesssim f\lesssim \omega_2$. Here the qubit population in the adiabatic leg of the drive (when the qubit is decoupled from both reservoirs) varies in time due to driving-induced coherent oscillations.
The analysis at even higher frequencies may not be rigorous as the drive frequency can exceed the inverse bath correlation time.
Considering the Otto refrigerator based on several qubits one can demonstrate that the correlation of the noises affecting both qubits separately can only diminish the efficiency of the refrigerator \cite{Karimi2017}.

Recently the quantum heat valve effect\footnote{\rev{This effect is based on the photon-mediated heat transfer quantization discussed in \cite{Meschke2006}. For more detailed discussion of heat quantization please see \cite{SM}.}}
has been reported \cite{Ronzani2018}, where the heat transport has been considered between
galvanically uncoupled systems, namely coplanar waveguide resonators, each terminated by a normal-metal mesoscopic resistor, only capacitively coupled via a superconducting transmon qubit.
It is shown that the reservoir-to-reservoir heat flux depends on the interplay between the qubit-resonator and the
resonator-reservoir couplings, yielding qualitatively dissimilar results in different coupling regimes.

\section{%Feedback systems - quantum Maxwell's Demon
FEEDBACK SYSTEMS - QUANTUM MAXWELL'S DEMON
}%Maxwell's Demons in superconducting quantum circuits
Over the past few years the idea of Maxwell's Demon presented in Sec.~\ref{Sec:MD} has been applied to quantum systems. %as well.
In principle an experiment analogous to the Szilard's Engine for electrons \cite{Koski2014_PRL,Koski2014_PNAS} can be implemented in a system consisting of a qubit, as proposed in Ref. \cite{Pekola2016PRB}.

The key element in this kind of a quantum MD is the type of measurement and feedback.
First, to avoid additional thermodynamic costs one has to perform a non-demolition quantum measurement of the qubit state such as the measurement of a qubit energy itself or its curvature.
Second, unlike the classical case in a qubit system there is an avoided level crossing in the vicinity of the degeneracy point (cf. the black lines in the insets of Fig.~\ref{Fig:SE_a_P(W)_b_Sagawa_Ueda}(a)).
It is this level anti-crossing which causes the change of the conditional feedback control.
Indeed, if the qubit is found in an excited state after the measurement, one has to apply additional $\pi$-pulse to extract the energy and drive the system to the ground state. Only after this one can change the control parameter detuning the qubit away from the degeneracy point.
Third, by optimizing the protocol of the control parameter change one can avoid additional excitations of the qubit due to Landau-Zener transitions \cite{Pekola2016PRB}.

Different experimental realizations have been reported recently \cite{Cottet2017,Masuyama2017,Naghiloo2018}.
However, direct measurement of heat extracted from the bath is still elusive and demanding.

\section{DISCLOSURE STATEMENT}
The authors are not aware of any affiliations, memberships, funding, or financial holdings that might be perceived as affecting the objectivity of this review.

\section{ACKNOWLEDGMENTS}
We thank
T.~Ala-Nissila,
D.~V.~Averin,
M.~Arzeo,
J.~Bergli,
M.~Campisi,
L.~Casparis,
Y.-C.~Chang,
C.~D.~Chen,
T.~Faivre,
R.~Fazio,
A.~V.~Feshchenko,
C.~Flindt,
Yu.~M.~Galperin,
S.~Gasparinetti,
F.~Giazotto,
D.~S.~Golubev,
W.~Guichard,
T.~T.~Heikkil\"a,
F.~W.~J.~Hekking,
M.~Helle,
F.~J\"ulicher,
B.~Karimi,
J.~V.~Koski,
V.~E.~Kravtsov,
A.~Kupiainen,
A.~Kutvonen,
M.~M.~Leivo,
A.~Luukanen,
V.~F.~Maisi,
D.~Maradan,
Y.~Masuyama,
M.~Meschke,
M.~M{\"o}tt{\"o}nen,
P.~Muratore-Ginanneschi,
Y.~Nakamura,
I.~Neri,
A.~O.~Niskanen,
M.~Palma,
J.~T.~Peltonen,
M.~Ribezzi-Crivellari,
F.~Ritort,
\'E.~Rold\'an,
A.~Ronzani,
T.~Sagawa,
O.-P.~Saira,
A.~M.~Savin,
K.~Schwieger,
A.~Shnirman,
J.~Senior,
S.~Singh,
P.~Solinas,
S.~Suomela,
T.~Tanttu,
A.~V.~Timofeev,
K.~L.~Viisanen,
Y.~Yoon,
and D.~M.~Zumb\"uhl
for insightful discussions and related collaborations.
J.~P.~P. acknowledges partial support through Academy of Finland, %though its LTQ CoE grant
Project Nos. 272218 % JPP's Academic grant
and 312057.% QTF CoE
, by European Research Council~(ERC) under the European Union's Horizon 2020 research
and innovation programme under grant agreement No. 742559 (SQH), % JPP's ERC
I.~M.~K. is supported by the German Research Foundation (DFG) Grant No. KH 425/1-1 and the Russian Foundation for Basic Research.

\bibliographystyle{ar-style4}
\bibliography{AR_MD_review}

\begin{thebibliography}{122}
\expandafter\ifx\csname natexlab\endcsname\relax\def\natexlab#1{#1}\fi

\bibitem{bochkov1977general}
Bochkov G, Kuzovlev YE. 1977.
\textit{Zh. Eksp. Teor. Fiz} 72:238--243

\bibitem{bochkov1981nonlinear}
Bochkov G, Kuzovlev YE. 1981.
\textit{Phys. A: Stat. Mech. Appl.} 106:443--479

\bibitem{Evans1993}
Evans DJ, Cohen EGD, Morriss GP. 1993.
\textit{Phys. Rev. Lett.} 71:2401--2404

\bibitem{Gallavotti1995}
Gallavotti G, Cohen EGD. 1995.
\textit{Phys. Rev. Lett.} 74:2694

\bibitem{Jarzynski1997}
Jarzynski C. 1997.
\textit{Phys. Rev. Lett.} 78:2690--2693

\bibitem{Crooks1999}
Crooks GE. 1999.
\textit{Phys. Rev. E} 60:2721

\bibitem{Seifert2005}
Seifert U. 2005.
\textit{Phys. Rev. Lett.} 95:040602

\bibitem{Seifert2012_RepProgPhys}
Seifert U. 2012.
\textit{Rep. Prog. Phys.} 75:126001

\bibitem{Sagawa2008}
Sagawa T, Ueda M. 2008.
\textit{Phys. Rev. Lett.} 100:080403

\bibitem{Sagawa2010}
Sagawa T, Ueda M. 2010.
\textit{Phys. Rev. Lett.} 104:090602

\bibitem{Toyabe2010}
Toyabe S, Sagawa T, Ueda M, Muneyuki E, Sano M. 2010.
\textit{Nat. Phys.} 6:988--992

\bibitem{Saira2012}
Saira OP, Yoon Y, Tanttu T, M\"ott\"onen M, Averin DV, Pekola JP. 2012.
\textit{Phys. Rev. Lett.} 109:180601

\bibitem{Koski2013}
Koski JV, Sagawa T, Saira OP, Yoon Y, Kutvonen A, et~al. 2013.
\textit{Nat. Phys.} 9:644

\bibitem{Berut2012}
B\'{e}rut A, Arakelyan A, Petrosyan A, Ciliberto S, Dillenschneider R, Lutz E.
  2012.
\textit{Nature} 483:187

\bibitem{Orlov2012}
Orlov AO, Lent CS, Thorpe CC, Boechler GP, Snider GL. 2012.
\textit{Japan. J. Appl. Phys.} 51:06FE10

\bibitem{Jun2014}
Jun Y, Gavrilov M, Bechhoefer J. 2014.
\textit{Phys. Rev. Lett.} 113:190601

\bibitem{Koski2014_PRL}
Koski JV, Maisi VF, Sagawa T, Pekola JP. 2014{\natexlab{a}}.
\textit{Phys. Rev. Lett.} 113:030601

\bibitem{Koski2014_PNAS}
Koski JV, Maisi VF, Pekola JP, Averin DV. 2014{\natexlab{b}}.
\textit{Proc. Nat. Acad. Sci.} 111:13786--13789

\bibitem{Roldan2014}
Rold\'an E, Mart\'inez IA, Parrondo JMR, Petrov D. 2014.
\textit{Nat. Phys.} 10:457--461

\bibitem{Koski2015}
Koski JV, Kutvonen A, Khaymovich IM, Ala-Nissil{\"a} T, Pekola JP. 2015.
\textit{Phys. Rev. Lett.} 115:260602

\bibitem{Khaymovich2015}
Khaymovich IM, Koski JV, Saira OP, Kravtsov VE, Pekola JP. 2015.
\textit{Nat. Comm.} 6:7010

\bibitem{Pekola2015NatPhys}
Pekola JP. 2015.
\textit{Nat. Phys.} 11:118

\bibitem{Hong2016}
Hong J, Lambson B, Dhuey S, Bokor J. 2016.
\textit{Science Adv.} 2:e1501492

\bibitem{Vidrighin2016}
Vidrighin MD, Dahlsten O, Barbieri M, Kim MS, Vedral V, Walmsley IA. 2016.
\textit{Phys. Rev. Lett.} 116:050401

\bibitem{Gavrilov2017}
Gavrilov M. 2017.
Erasure without work in an asymmetric, double-well potential. In
  \textit{Experiments on the Thermodynamics of Information Processing}.
  Springer,  83--96

\bibitem{Ribezzi}
Ribezzi-Crivellari M, Ritort F, et~al. 2015.
private communication

\bibitem{Collin2005}
Collin D, Ritort F, Jarzynski C, Smith SB, Tinoco~Jr I, Bustamante C. 2005.
\textit{Nature} 437:231

\bibitem{Alemany2011}
Alemany A, Ribezzi M, Ritort F. 2011.
Recent progress in fluctuation theorems and free energy recovery. In
  \textit{AIP Conf. Proc.}, vol. 1332. AIP

\bibitem{Alemany2015}
Alemany A, Ribezzi-Crivellari M, Ritort F. 2015.
\textit{New J. Phys.} 17:075009

\bibitem{Rowell1976}
Rowell JM, Tsui DC. 1976.
\textit{Phys. Rev. B} 14:2456--2463

\bibitem{Feshchenko2015}
Feshchenko AV, Casparis L, Khaymovich IM, Maradan D, Saira OP, et~al. 2015.
\textit{Phys. Rev. Applied} 4:034001

\bibitem{Nahum1994}
Nahum M, Eiles TM, Martinis JM. 1994.
\textit{Appl. Phys. Lett.} 65:3123--3125

\bibitem{Leivo1996}
Leivo MM, Pekola JP, Averin DV. 1996.
\textit{Appl. Phys. Lett.} 68:1996--1998

\bibitem{Clark2005}
Clark AM, Miller NA, Williams A, Ruggiero ST, Hilton GC, et~al. 2005.
\textit{Appl. Phys. Lett.} 86:173508

\bibitem{Giazotto2006}
Giazotto F, Heikkil\"a TT, Luukanen A, Savin AM, Pekola JP. 2006.
\textit{Rev. Mod. Phys.} 78:217--274

\bibitem{Lebowitz1999}
Lebowitz JL, Spohn H. 1999.
\textit{J. Stat. Phys.} 95:333--365

\bibitem{Shargel2009}
Shargel BH, Chou T. 2009.
\textit{J. Stat. Phys.} 137:165

\bibitem{Schuler2005}
Schuler S, Speck T, Tietz C, Wrachtrup J, Seifert U. 2005.
\textit{Phys. Rev. Lett.} 94:180602

\bibitem{Tietz2006}
Tietz C, Schuler S, Speck T, Seifert U, Wrachtrup J. 2006.
\textit{Phys. Rev. Lett.} 97:050602

\bibitem{Mandaiya2018}
Mandaiya A, Khaymovich IM. 2018.
Relations between long-time and finite-time fluctuation theorems in two-level
  system under periodic drive.
in preparation

\bibitem{Chetrite2011}
Chetrite R, Gupta S. 2011.
\textit{J. Stat. Phys.} 143:543

\bibitem{Neri2017}
Neri I, Rold\'an E, J\"ulicher F. 2017.
\textit{Phys. Rev. X} 7:011019

\bibitem{Pigolotti2017}
Pigolotti S, Neri I, Rold\'an E, J\"ulicher F. 2017.
\textit{Phys. Rev. Lett.} 119:140604

\bibitem{Fujisawa2006}
Fujisawa T, Hayashi T, Tomita R, Hirayama Y. 2006.
\textit{Science} 312:1634

\bibitem{Kung2012}
K\"ung B, R\"ossler C, Beck M, Marthaler M, Golubev DS, et~al. 2012.
\textit{Phys. Rev. X} 2:011001

\bibitem{Singh2017}
{Singh} S, {Rold{\'a}n} {\'E}, {Neri} I, {Khaymovich} IM, {Golubev} DS, et~al.
  2017.
\textit{\rm arXiv} :1712.01693

\bibitem{Leff2002}
Leff H, Rex AF. 2002.
Maxwell's demon 2 entropy, classical and quantum information, computing.
CRC Press

\bibitem{Maruyama2009}
Maruyama K, Nori F, Vedral V. 2009.
\textit{Rev. Mod. Phys.} 81:1

\bibitem{Parrondo2015}
Parrondo JM, Horowitz JM, Sagawa T. 2015.
\textit{Nat. Phys.} 11:131

\bibitem{Landauer1961}
Landauer R. 1961.
\textit{IBM J. Res. Develop.} 5:183--191

\bibitem{Landauer1988}
Landauer R. 1988.
\textit{Nature} 335:779--784

\bibitem{Bergli2013}
Bergli J, Galperin YM, Kopnin N. 2013.
\textit{Phys. Rev. E} 88:062139

\bibitem{Sordal2017}
S{\o}rdal V, Bergli J, Galperin Y. 2017.
\textit{Phys. Rev. E} 95:062129

\bibitem{Walldorf2017}
Walldorf N, Jauho AP, Kaasbjerg K. 2017.
\textit{Phys. Rev. B} 96:115415

\bibitem{Averin2017}
Averin DV, Pekola JP. 2017.
\textit{Phys. Stat. Sol. (b)} 254:1600677

\bibitem{Chida2015}
Chida K, Nishiguchi K, Yamahata G, Tanaka H, Fujiwara A. 2015.
\textit{Appl. Phys. Lett.} 107:073110

\bibitem{Wagner2016}
Wagner T, Strasberg P, Bayer JC, Rugeramigabo EP, Brandes T, Haug RJ. 2016.
\textit{Nat. Nanotech.} 12:218

\bibitem{Singh2016}
Singh S, Peltonen JT, Khaymovich IM, Koski JV, Flindt C, Pekola JP. 2016.
\textit{Phys. Rev. B} 94:241407

\bibitem{Strasberg2013}
Strasberg P, Schaller G, Brandes T, Esposito M. 2013.
\textit{Phys. Rev. Lett.} 110:040601

\bibitem{Horowitz2014}
Horowitz JM, Esposito M. 2014.
\textit{Phys. Rev. X} 4:031015

\bibitem{Shiraishi2015}
Shiraishi N, Ito S, Kawaguchi K, Sagawa T. 2015.
\textit{New J. Phys.} 17:045012

\bibitem{Sanchez2011}
S{\'a}nchez R, B{\"u}ttiker M. 2011.
\textit{Phys. Rev. B} 83:085428

\bibitem{Sanchez2012}
S{\'a}nchez R, B{\"u}ttiker M. 2012.
\textit{Europhys. Lett.} 100:47008

\bibitem{Thierschmann2015NatNano}
Thierschmann H, S{\'a}nchez R, Sothmann B, Arnold F, Heyn C, et~al.
  2015{\natexlab{a}}.
\textit{Nat. Nanotech.} 10:854--858

\bibitem{Thierschmann2015NJP}
Thierschmann H, Arnold F, Mitterm{\"u}ller M, Maier L, Heyn C, et~al.
  2015{\natexlab{b}}.
\textit{New J. Phys.} 17:113003

\bibitem{SM}
 2018.
Follow the Supplemental Materials link in the online version of this article or
  at http://www.annualreviews.org/."

\bibitem{Pekola2014}
Pekola JP, Koski JV, Averin DV. 2014.
\textit{Phys. Rev. B} 89:081309(R)

\bibitem{Feshchenko2014}
Feshchenko AV, Koski JV, Pekola JP. 2014.
\textit{Phys. Rev. B} 90:201407(R)

\bibitem{Esposito2009}
Esposito M, Harbola U, Mukamel S. 2009.
\textit{Rev. Mod. Phys.} 81:1665

\bibitem{Campisi2011}
Campisi M, H{\"a}nggi P, Talkner P. 2011.
\textit{Rev. Mod. Phys.} 83:771

\bibitem{Dalibard1992}
Dalibard J, Castin Y, M\o{}lmer K. 1992.
\textit{Phys. Rev. Lett.} 68:580--583

\bibitem{Hekking2013}
Hekking FWJ, Pekola JP. 2013.
\textit{Phys. Rev. Lett.} 111:093602

\bibitem{Horowitz2013}
Horowitz JM, Parrondo JMR. 2013.
\textit{New J. Phys.} 15:085028

\bibitem{Suomela2016}
Suomela S, Kutvonen A, Ala-Nissila T. 2016.
\textit{Phys. Rev. E} 93:062106

\bibitem{Pekola2016JLTP}
Pekola JP, Suomela S, Galperin YM. 2016.
\textit{J. Low Temp. Phys.} 184:1015--1029

\bibitem{Pekola2015PRE}
Pekola JP, Masuyama Y, Nakamura Y, Bergli J, Galperin YM. 2015.
\textit{Phys. Rev. E} 91:062109

\bibitem{Kupiainen2016}
Kupiainen A, Muratore-Ginanneschi P, Pekola J, Schwieger K. 2016.
\textit{Phys. Rev. E} 94:062127

\bibitem{Pekola2013}
Pekola JP, Solinas P, Shnirman A, Averin DV. 2013.
\textit{New J. Phys.} 15:115006

\bibitem{Partanen2017}
{Partanen} M, {Yen Tan} K, {Masuda} S, {Govenius} J, {Lake} RE, et~al. 2017.
\textit{\rm arXiv} :1712.10256

\bibitem{Ronzani2018}
{Ronzani} A, {Karimi} B, {Senior} J, {Chang} YC, {Peltonen} JT, et~al. 2018.
\textit{\rm arXiv} :1801.09312

\bibitem{Gasparinetti2015}
Gasparinetti S, Viisanen KL, Saira OP, Faivre T, Arzeo M, et~al. 2015.
\textit{Phys. Rev. Applied} 3:014007

\bibitem{Zgirski2017}
Zgirski M, Foltyn M, Savin A, Meschke M, Pekola J. 2017.
\textit{\rm arXiv} :1704.04762

\bibitem{Wang2018}
Wang L, Saira OP, Pekola J. 2018.
\textit{Appl. Phys. Lett.} 112:013105

\bibitem{Kosloff2014}
Kosloff R, Levy A. 2014.
\textit{Annu. Rev. Phys. Chem.} 65:365--393

\bibitem{Alicki1979}
Alicki R. 1979.
\textit{J. Phys. A: Math. Gen.} 12:L103

\bibitem{Campisi2016}
Campisi M, Fazio R. 2016.
\textit{Nat. Comm.} 7:11895

\bibitem{Hofer2016_PRB93}
Hofer PP, Souquet JR, Clerk AA. 2016.
\textit{Phys. Rev. B} 93:041418

\bibitem{Scully2003}
Scully MO, Zubairy MS, Agarwal GS, Walther H. 2003.
\textit{Science} 299:862--864

\bibitem{Quan2007}
Quan HT, Liu Yx, Sun CP, Nori F. 2007.
\textit{Phys. Rev. E} 76:031105

\bibitem{Marchegiani2016}
Marchegiani G, Virtanen P, Giazotto F, Campisi M. 2016.
\textit{Phys. Rev. Applied} 6:054014

\bibitem{Uzdin2015}
Uzdin R, Levy A, Kosloff R. 2015.
\textit{Phys. Rev. X} 5:031044

\bibitem{Campisi2017}
Campisi M, Pekola J, Fazio R. 2017.
\textit{New J. Phys.} 19:053027

\bibitem{Abah2016}
Abah O, Lutz E. 2016.
\textit{Europhys. Lett.} 113:60002

\bibitem{Brandner2016}
Brandner K, Seifert U. 2016.
\textit{Phys. Rev. E} 93:062134

\bibitem{Niskanen2007}
Niskanen AO, Nakamura Y, Pekola JP. 2007.
\textit{Phys. Rev. B} 76:174523

\bibitem{Hofer2016_PRB94}
Hofer PP, Perarnau-Llobet M, Brask JB, Silva R, Huber M, Brunner N. 2016.
\textit{Phys. Rev. B} 94:235420

\bibitem{Karimi2016}
Karimi B, Pekola JP. 2016.
\textit{Phys. Rev. B} 94:184503

\bibitem{Karimi2017}
Karimi B, Pekola JP. 2017.
\textit{Phys. Rev. B} 96:115408

\bibitem{Rossnagel2016}
Ro{\ss}nagel J, Dawkins ST, Tolazzi KN, Abah O, Lutz E, et~al. 2016.
\textit{Science} 352:325--329

\bibitem{Campisi2014}
Campisi M. 2014.
\textit{J. Phys. A: Math. Theor.} 47:245001

\bibitem{Ptaszynski}
Ptaszynski K. 2018.
\textit{\rm arXiv} :1805.11301

\bibitem{Breuer2002}
Breuer H, Petruccione F. 2002.
The theory of open quantum systems.
Oxford University Press

\bibitem{Pekola2016PRB}
Pekola JP, Golubev DS, Averin DV. 2016.
\textit{Phys. Rev. B} 93:024501

\bibitem{Meschke2006}
Meschke M, Guichard W, Pekola JP. 2006.
\textit{Nature} 444:187

\bibitem{Cottet2017}
Cottet N, Jezouin S, Bretheau L, Campagne-Ibarcq P, Ficheux Q, et~al. 2017.
\textit{Proc. Nat. Acad. Sci.} 114:7561--7564

\bibitem{Masuyama2017}
{Masuyama} Y, {Funo} K, {Murashita} Y, {Noguchi} A, {Kono} S, et~al. 2017.
\textit{\rm arXiv} :1709.00548

\bibitem{Naghiloo2018}
Naghiloo M, Alonso J, Romito A, Lutz E, Murch K. 2018.
\textit{\rm arXiv} :1802.07205

\bibitem{Landauer1957}
Landauer R. 1957.
\textit{IBM J. Res. Develop.} 1:223--231

\bibitem{vanWees1988}
van Wees BJ, van Houten H, Beenakker CWJ, Williamson JG, Kouwenhoven LP, et~al.
  1988.
\textit{Phys. Rev. Lett.} 60:848--850

\bibitem{Wharam1988}
Wharam DA, Thornton TJ, Newbury R, Pepper M, Ahmed H, et~al. 1988.
\textit{J. Phys. C: Solid Stat. Phys.} 21:L209

\bibitem{Pendry1983}
Pendry JB. 1983.
\textit{J. Phys. A: Math. Gen.} 16:2161

\bibitem{Rego1999}
Rego LGC, Kirczenow G. 1999.
\textit{Phys. Rev. B} 59:13080--13086

\bibitem{Pendry1999}
Pendry JB. 1999.
\textit{J. Phys.: Cond. Matt.} 11:6621

\bibitem{Schmidt2004}
Schmidt DR, Schoelkopf RJ, Cleland AN. 2004.
\textit{Phys. Rev. Lett.} 93:045901

\bibitem{Schwab2000}
Schwab K, Henriksen EA, Worlock JM, Roukes ML. 2000.
\textit{Nature} 404:974

\bibitem{Yung2002}
Yung CS, Schmidt DR, Cleland AN. 2002.
\textit{Appl. Phys. Lett.} 81:31--33

\bibitem{Chiatti2006}
Chiatti O, Nicholls JT, Proskuryakov YY, Lumpkin N, Farrer I, Ritchie DA. 2006.
\textit{Phys. Rev. Lett.} 97:056601

\bibitem{Jezouin2013}
Jezouin S, Parmentier FD, Anthore A, Gennser U, Cavanna A, et~al. 2013.
\textit{Science} 342:601--604

\bibitem{Timofeev2009}
Timofeev AV, Helle M, Meschke M, M\"ott\"onen M, Pekola JP. 2009.
\textit{Phys. Rev. Lett.} 102:200801

\bibitem{Partanen2016}
Partanen M, Tan KY, Govenius J, Lake RE, M{\"o}kel{\"o} MK, et~al. 2016.
\textit{Nat. Phys.} 12:460

\bibitem{Ciliberto2013}
Ciliberto S, Imparato A, Naert A, Tanase M. 2013.
\textit{Phys. Rev. Lett.} 110:180601

\bibitem{Golubev2015}
Golubev DS, Pekola JP. 2015.
\textit{Phys. Rev. B} 92:085412

\end{thebibliography}

\appendix

\section{Schematics of Autonomous Maxwell's Demon}

\subsection{Schematics}
\begin{figure}
\includegraphics[width=0.6\textwidth]{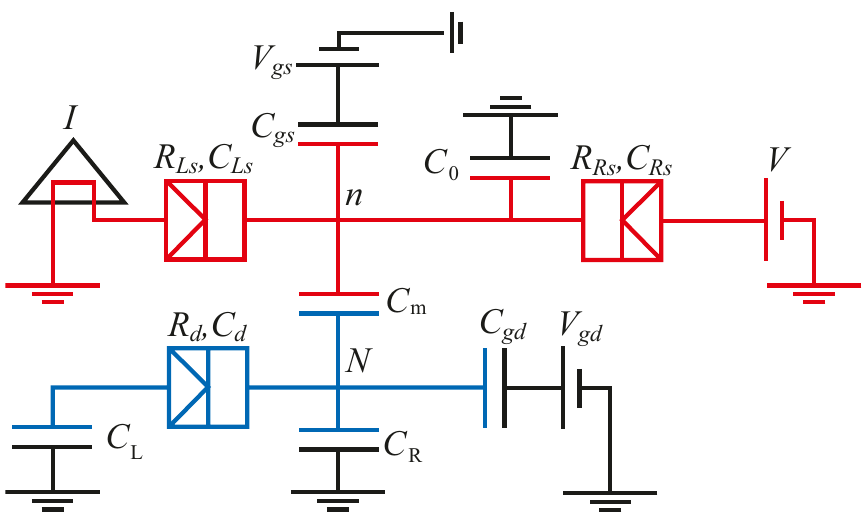}
\caption{
Schematic representation of the measurement setup of the autonomous Maxwell's Demon experiments.
The system single-electron transistor (shown in red) is biased with the voltage $V$ and capacitively coupled to
the detector single-electron box formed by two capacitively coupled islands (shown in blue).
The current through the system as well as the temperatures of System and Demon electrodes (not shown) are measured in the experiments.
All connections, voltages, resistances, capacitances and charges (in an electron charge) are represented in the figure.
}
\label{SM_Fig:AutoMD_schematics}
\end{figure}

In the experiment on the autonomous Maxwell's Demon (MD), the Demon is made by an unbiased SET (equivalent to SEB) with $N$ excess electrons on the island, capacitively coupled to the System SET (see Supplemental Figure~\ref{SM_Fig:AutoMD_schematics}). This configuration resembles theoretical proposals of heat-to-current convertors on quantum dots \cite{Sanchez2011,Sanchez2012} realized in several experiments up to now \cite{Thierschmann2015NatNano,Thierschmann2015NJP}.
The resistance of the detector junctions $R_d$ determining its operating frequency by tunneling rates
is made much smaller than the one in the System $R_s$, $R_d\ll R_s$.
Here for simplicity we consider both detector (Demon) and System with equal left and right tunnel junctions, i.e., $R_{Ls}=R_{Rs}\equiv R_s$.
The energy of the supersystem of two coupled single-electronic devices
\begin{equation}
U(n,N,n_g,N_g)=E_{C,s}(n-n_g)^2 + E_{C,d}(N-N_g)^2 + 2J(n-n_g)(N-N_g).
\end{equation}
The corresponding Coulomb energies
\begin{gather}
E_{C,s} = e^2 C_{\Sigma,d}/2(C_{\Sigma,s}C_{\Sigma,d}-C_m^2), \\
E_{C,d}=e^2 C_{\Sigma,s}/2(C_{\Sigma,s}C_{\Sigma,d}-C_m^2), \\
J=e^2 C_{m}/2(C_{\Sigma,s}C_{\Sigma,d}-C_m^2)
\end{gather}
depend on the total capacitances of the System $C_{\Sigma,s} = C_{Ls}+C_{Rs}+C_{gs}+C_0+C_m$ and the Demon $C_{\Sigma,d} = C_{d}+C_{gd}+C_R+C_m$ islands and on their mutual capacitance $C_m$.
The polarization charges $n_g = C_{gs} V_{gs}/e$ and $N_g = C_{gd} V_{gd}/e$ are governed by the gate voltages $V_{gs}$, $V_{gd}$.

%=====
\subsection{General expressions for the dissipated heat and mutual information under feedback}
Following the main text we consider the simplest case of the feedback control with
maximal coupling between System and Demon, $N_g = 1/2$,
and maximal current flowing through the System, $n_g=1/2$ and assume temperatures to be quite small $k_B T_L, k_B T_R, k_B T_{det}\ll E_{C,s},E_{C,d}$ comparing to the above mentioned charging energies.
In this case $n$ and $N$ are limited to two possible values $0$ and $1$
and the electrostatic energy of the overall supersystem depends on these variables only via the last term
%give just a constant offset as $N,n = \overline{0,1}$ and the
\begin{gather}
\delta U (n,N) = U(n,N,1/2,1/2)-E_{C,s}/4 - E_{C,d}/4 = J(2n-1)(2N-1)/2 \ .
\end{gather}
Considering for simplicity the equal System junctions and temperatures $T_L=T_R\equiv T_s$, one can prove that steady state probabilities
$P_{n,N}$ to realize a state of ground-state $P_{0,1}$, $P_{1,0}$ and excited-state $P_{0,0}$, $P_{1,1}$ doublets take the form
\begin{subequations}
\begin{align}
P_{0,1} = P_{1,0}\equiv P_g/2 \ ,\\
P_{0,0} = P_{1,1}\equiv P_e/2 \ ,
\end{align}
\end{subequations}
with $P_g = \Gamma_+/(\Gamma_+ +\Gamma_-)$, $P_e = \Gamma_-/(\Gamma_+ +\Gamma_-)$,
and the upper (lower) sign in $\Gamma_\pm = \Gamma_s(\pm J_+)+\Gamma_s(\pm J_-)+\Gamma_d(\pm J)$
corresponds to the tunneling to the ground (excited) state.
The expressions for System (Demon) tunneling rates $\Gamma_{s (d)}$ are given in the main text.

As shown in the main text the ideal heat generation rates in the System and in the Demon take the forms
$\dot Q_s^{ideal} = -(2J/e-V)I$ and $\dot Q_d^{ideal} = 2 J I/e$ provided the relaxation rate to the ground state is mostly dominated by the tunneling in the detector $\Gamma_+\simeq \Gamma_d(J)$ as well as the tunneling in the System mostly brings the total supersystem to the excited state $\Gamma_- \simeq \Gamma_s(-J_-)\ll \Gamma_+$.

In the general case, one should take into account all possible processes and get the following expression for the average System heat rate
\begin{multline}\label{e18_Auto_MD_Qs}
\dot Q_s = -P_g \left[J_- \Gamma_s(-J_-)+J_+ \Gamma_s(-J_+)\right]+P_e\left[J_- \Gamma_s(J_-)+J_+ \Gamma_s(J_+)\right]\\
\simeq
-\frac{\Gamma_s(-J_-)}{\Gamma_d(J)} P_g\left[J_- \Gamma_d(J) - J_- \Gamma_s(J_-)-J_+ \Gamma_s(J_+)\right] \ ,
\end{multline}
which can be negative provided $J-eV/2>0$, $P_e\simeq P_g \Gamma_s(-J_-)/\Gamma_d(J)$ and $\Gamma_d(J)\gg \Gamma_s(J_\pm)$.
The strict condition on the resistance is given by $d^2 \dot Q_s/ d V^2|_{V=0}<0$
and reads as
\begin{gather}
\frac{J}{4k_B T_s} \coth\left(\frac{J}{2 k_B T_s}\right)>\left(1+\frac{R_d}{R_s}\right)
\end{gather}
for $T_s = T_{det}$, see \cite{Koski2015} for details.
Note that in any case the average System current is still flowing towards bias
\begin{gather}
2I/e = P_g \left[\Gamma_s(-J_-)-\Gamma_s(-J_+)\right]+P_e\left[\Gamma_s(J_+)-\Gamma_s(J_-)\right]>0 .
\end{gather}
The heat dissipation satisfying the Joule's law is taken by the heat flow rate in the Demon
\begin{gather}\label{e19_Auto_MD_Qd}
\dot Q_d=I V-\dot Q_s = -J\Gamma_d(-J) P_g + J\Gamma_d(J) P_e \simeq J P_g [\Gamma_s(-J_-)-\Gamma_d(-J)]>0 \ .
\end{gather}

\begin{figure}
\includegraphics[width=0.6\textwidth]{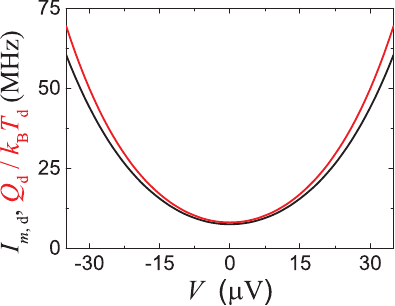}
\caption{
Numerical comparison between the heat dissipation rate in the Maxwell's Demon normalized to its temperature and the rate of the mutual information gained by the Demon. This plot demonstrates that the two quantities match within $15~\%$ of error.}
\label{SM_Fig:AutoMD_I_M}
\end{figure}

%=====
As at $N_g = n_g = 1/2$ the marginal probabilities are identical $P_{n} = P_{N} = 1/2$, the mutual information
\begin{gather}\label{e11_mutual_info}
I_M (n| m) = \ln P_{n, m} - \ln P_{n} - \ln P_{m} \ ,
\end{gather}
changes only due to change of $P_{n,N}$ between $P_g/2$ and $P_e/2$.
The tunneling in the Demon gives the following contribution to the mutual information rate
\begin{gather}\label{e20_Auto_MD_I_M,d}
\dot I_{M,d} = \left[\Gamma_d(J) P_e - \Gamma_d(-J) P_g\right]\ln(P_g/P_e)\simeq \left[\Gamma_s(-J_-) - \Gamma_d(-J)\right]P_g\ln(P_g/P_e)>0 \ ,
\end{gather}
which is positive within the same assumptions as taken for \eqref{e18_Auto_MD_Qs} and \eqref{e19_Auto_MD_Qd}.
As the process is cyclic the average mutual information is conserved and the contribution to its rate governed by the System tunneling is $\dot I_{M,s}=-\dot I_{M,d}$.
According to \cite{Horowitz2014} the above mentioned mutual information rates place an upper bound to
the System heat absorbtion rate $-\dot Q_s \leq k_B T_s \dot I_{M,d}$ and the lower bound to the heat rate generated in MD $\dot Q_d \geq k_B T_d \dot I_{M,d}$.
The latter inequality is saturated in the limit $R_d\ll R_s$ as in this case the probability distribution has a thermal form
$\ln(P_g/P_e) = 2J/k_B T_d$ and the r.h.s. of Eqs.~\eqref{e19_Auto_MD_Qd}, \eqref{e20_Auto_MD_I_M,d} coincide.
In the experiment the above mentioned inequality $\dot{Q}_d \geq k_B T_d I_{M,d}$ differs from equality only by $15$~\%, Supplemental Figure~\ref{SM_Fig:AutoMD_I_M},
which verifies the theoretical predictions for MD thermodynamics.

%==============================================
\section{Photon-mediated heat transport}
The concept of charge conductance quantization \cite{Landauer1957,vanWees1988,Wharam1988} is well-known and established in condensed matter physics. The corresponding concept of the quantization of thermal conductance $G_{Th}$ with a heat conductance quantum $G_Q = \pi k_B^2 T/6\hbar$
suggested in \cite{Pendry1983} is less known, however it is more general \cite{Rego1999,Pendry1999,Schmidt2004} as it works for
excitations regardless on their statistics (for experiments verifying this thermal conductance quantization please see
\cite{Schwab2000,Yung2002} for phonons,
\cite{Chiatti2006,Jezouin2013} for electrons, and
\cite{Meschke2006,Timofeev2009,Partanen2016} for photons).

\begin{figure}
\includegraphics[width=0.8\textwidth]{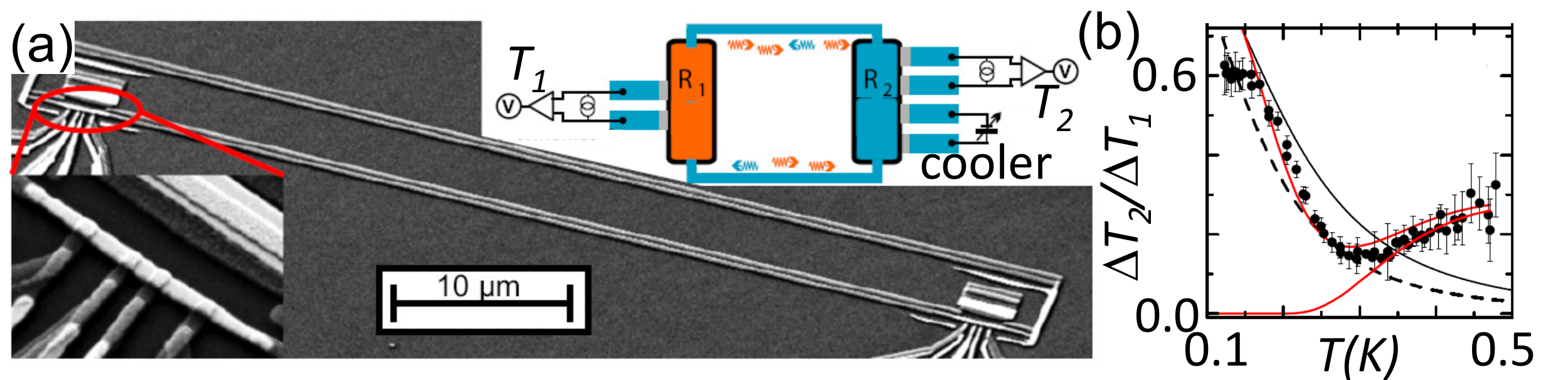}
\caption{(a)~Electron micrograph of the sample
consisting of two normal metal (AuPd) islands playing the role of resistors at a $50~\mu$m distance from each other,
connected with superconducting (Al) lines into a loop to enable remote photon-mediated heat transport.
(lower left inset) Zoomed atomic force microscopy image of the island.
The four NIS junctions, contacting each island in the middle part, are used
to perturb and to measure the island temperature.
(upper right inset) Equivalent electrical circuit of the structure.
(b)~Relative temperature changes of the second resistor at $T=120-500$~mK, when the first one is heated up $\Delta T_1 = T_1-T>0$.
The data (black symbols) matches with the theoretical predictions for
the full quantum conductance $G_{Th}=G_Q$ and vanishing quasiparticle conductance through the superconductor (red lines) and
deviating from the ones with smaller $G_{Th}$.
Panel~(b) is adapted from \cite{Timofeev2009}.
}
\label{Fig:Photon_Heat_Transport}
\end{figure}

In this section we address mostly the photon-mediated mechanism of heat transport and focus on experimental achievements in this direction.
The experimental verification of the photon-mediated thermal conductance quantization has been reported for the first time in \cite{Meschke2006}.
There the authors realized the system of two resistors $R_1$ and $R_2$ playing the role of thermal reservoirs put in general at different temperatures $T_1$ and $T_2$, respectively. These reservoirs are galvanically connected via superconducting circuits allowing dissipationless charge transport and avoiding quasiparticle heat transport Supplemental Figure~\ref{Fig:Photon_Heat_Transport}(a).
The phonon heat transport has been also diminished by putting the system to low enough temperatures.
The virtually totally reactive impedance $Z(\omega)$ of the superconducting circuits made of two DC-SQUID devices has been tuned by tiny external magnetic field to separate the effect of photon-mediated heat transfer from other mechanisms.
The variations of the impedance of the galvanic connection via changing of magnetic flux varies the photon heat transfer leading to the oscillations of the resistor temperatures.
In agreement with the theoretical prediction the amplitude of these oscillations decays with the bath temperature $T$ close to the equilibrium and demonstrates non-monotonic behavior in the regime of heating of one of resistors.

In the follow-ups of \cite{Meschke2006} there has been demonstrated several improvements both
in the amplitude of photon-mediated heat conductance via optimization of the electronic circuit matching \cite{Timofeev2009}
and in photon-path distances between electronic systems \cite{Partanen2016}.
These proof-of-concept experiments utilize photon-mediated electronic refrigerators based on the following principle.
Using the similar setup as in \cite{Meschke2006} the authors of \cite{Timofeev2009,Partanen2016} cooled down one of normal resistors
via tunnel NIS junctions (for details see the electronic refrigeration section in the main text) and monitored the electronic temperatures of both resistors (for details see the electronic thermometry section of the main text), see the upper inset of Supplemental Figure~\ref{Fig:Photon_Heat_Transport}(a).
Due to the matching of electronic circuits the refrigeration of the distant resistor has been shown \cite{Timofeev2009} to be limited by the heat conductance quantum $G_Q$ (see the corresponding plot in Supplemental Figure~\ref{Fig:Photon_Heat_Transport}(b)).
The paper \cite{Partanen2016} demonstrates the absence of a principal upper bound on photon-path distance between reservoirs.

The fluctuations of the heat transfer between two resistors
kept at different temperature capacitively coupled to each other
have been addressed both theoretically and experimentally in \cite{Ciliberto2013}.
In the classical limit of rather large temperatures authors determined the out-of-equilibrium heat variance
as function of the temperature difference, the heat flux.
The theoretical paper \cite{Golubev2015} generalizes this result to any reactive impedance $Z(\omega)$ of the connecting lines and applies both to the classical and quantum limits of temperatures.
This work shows that fluctuations of heat transferred between the resistors are determined by random scattering of photons
on an effective barrier with frequency dependent transmission probability.

Recently the quantum heat valve effect has been reported \cite{Ronzani2018}, where the heat transport has been considered between
galvanically uncoupled systems, namely coplanar waveguide resonators, each terminated by a normal-metal mesoscopic resistor, only capacitively coupled via a superconducting transmon qubit.
It is shown that the reservoir-to-reservoir heat flux depends on the interplay between the qubit-resonator and the
resonator-reservoir couplings, yielding qualitatively dissimilar results in different coupling regimes.

\end{document}